\documentclass[iop,tighten,twocolumn,apj,floatfix]{emulateapj}

\usepackage{multirow}
\usepackage{amssymb}
\usepackage{graphicx}
\usepackage{amsmath}
\usepackage{cprotect} 
\usepackage{paralist}
\usepackage{bm}
\usepackage[breaklinks,colorlinks,citecolor=blue]{hyperref} 
\usepackage[caption=false]{subfig}

\begin{document}

\title{The Distribution of Ultra-Diffuse and Ultra-Compact Galaxies in the Frontier Fields}

\author{Steven~R.~Janssens\altaffilmark{1}, Roberto~Abraham\altaffilmark{1},
Jean~Brodie\altaffilmark{2}, Duncan~A.~Forbes\altaffilmark{3}, and
Aaron~J.~Romanowsky\altaffilmark{2,4}}

\altaffiltext{1}{Department of Astronomy and Astrophysics, University of
Toronto, 50 St. George Street, Toronto, ON, Canada M5S 3H4}
\altaffiltext{2}{University of California Observatories, 1156 High Street,
Santa Cruz, CA 95064, USA}
\altaffiltext{3}{Centre for Astrophysics and Supercomputing, Swinburne
University, Hawthorn VIC 3122, Australia}
\altaffiltext{4}{Department of Physics and Astronomy, San Jos\'e State
University, One Washington Square, San Jose, CA 95192, USA}

\email{janssens@astro.utoronto.ca}

\shorttitle{UDGs in the Frontier Fields}
\shortauthors{Janssens et al.}

\begin{abstract}
Large low surface brightness galaxies have recently been found to be abundant
in nearby galaxy clusters. In this paper, we investigate these
ultra-diffuse galaxies (UDGs) in the six \textit{Hubble} Frontier Fields
galaxy clusters: Abell~2744, MACSJ0416.1$-$2403,
MACSJ0717.5$+$3745, MACSJ1149.5$+$2223, Abell~S1063 and Abell~370.  These
are the most massive ($1$--$3 \times 10^{15}~M_\odot$) and distant ($0.308
< z < 0.545$) systems in which this class of galaxy has yet been
discovered. We estimate that the clusters host of the order of ${\sim}$200--1400 UDGs
inside the virial radius ($R_{200}$), consistent with the UDG abundance
halo-mass relation found in the local universe, and suggests that UDGs may
be formed in clusters. Within each cluster, however, we find that UDGs are
not evenly distributed. Instead their projected spatial distributions are
lopsided, and they are deficient in the regions of highest mass density as
traced by gravitational lensing. While the deficiency of UDGs in central
regions is not surprising, the lopsidedness is puzzling. The UDGs, and their
lopsided spatial distributions, may be
associated with known substructures late in their infall into the
clusters, meaning that we find evidence both for formation of UDGs in clusters
and for UDGs falling into clusters. We also investigate the ultra-compact
dwarfs (UCDs) residing in the clusters, and find the spatial distributions
of UDGs and UCDs appear anti-correlated. Around 15\% of UDGs exhibit
either compact nuclei or nearby point sources. Taken together, these
observations provide additional evidence for a picture in which at least some
UDGs are destroyed in dense cluster environments and leave behind a
residue of UCDs.
\end{abstract}


\section{Introduction}

Large low surface brightness galaxies in galaxy clusters were first reported
by \cite{vcc3} who noted the existence of ``very-large-size,
low-surface-brightness dwarfs" in the Virgo cluster.
They chose not to introduce a new morphological designation for these
galaxies, since apart from their large sizes ($\sim$10 kpc in diameter), they
resemble otherwise normal dwarfs or dwarf irregulars \citep{vcc2}.
Similarly, \cite{impey1988} found an additional 27 examples in
Virgo, and similar large low surface brightness galaxies were found in Fornax
\citep{ferguson1988, bothun1991}.
Related objects were also found in lower density environments, such as the low
surface brightness galaxies described by \cite{dalcanton1997}, and F8D1 in the
M81 group \citep{caldwell1998}.

The interest in large low-surface-brightness galaxies has recently been
reignited by the discovery of the enormous abundance of such `extreme' low
surface brightness systems in the richest environments, most notably the
discovery of very large numbers of such systems in the Coma cluster
\citep[see][]{vandokkum2015a, koda2015}, coupled with the remarkable properties of
the small number of such objects that have been investigated in detail by
follow-up investigations \citep[e.g.][]{beasley2016a, vandokkum2016,
vandokkum2018a, danieli2019a, vandokkum2019a, martin-navarro2019}.
At least a subset of such objects appear to host very extensive globular
cluster systems and exhibit anomalous dynamical mass-to-light ratios (described
in greater detail below).
As a result of this, and following a suggestion by \cite{vandokkum2015a}, very
large low-surface brightness galaxies in clusters have come to be known as
`ultra-diffuse galaxies' (UDGs). 

There is no universally accepted definition of a UDG in the literature.
Slight differences arise from the variety of instruments and techniques used
\citep{martin2019}.
However, they are typically defined in morphological terms as very extended
stellar systems with large effective radii ($R_e \gtrsim 1.5~\mathrm{kpc}$),
low S\'{e}rsic indices ($n \lesssim 1.5$), and characteristic surface
brightnesses fainter than ${\sim}24~\mathrm{mag}~\mathrm{arcsec}^{-2}$, i.e.\
roughly Milky Way sized with 1/100--1/1000 the stellar mass.
\cite{vandokkum2015a} chose to adopt the central surface brightness of UDGs as
the characteristic surface brightness of the class, and required this to be
fainter than $\mu_{0,g} \approx 24~\mathrm{mag}~\mathrm{arcsec}^{-2}$.
On the other hand, \cite{koda2015} adopted the mean surface-brightness inside
$R_e$ in the $R$-band as the characteristic surface brightness, and defined
their sample using a cut of $\langle\mu\rangle_{e,R} >
24~\mathrm{mag}~\mathrm{arcsec}^{-2}$.
Concentrated, likely background, systems were removed by requiring
$\langle\mu\rangle_e$ to not significantly deviate from the surface brightness
at $R_e$.
Similarly, \cite{vdb2016} used an $r$-band surface brightness cut of
$\langle\mu\rangle_{e,r} \geq 24~\mathrm{mag}~\mathrm{arcsec}^{-2}$ and removed
concentrated systems with a S\'{e}rsic index cut of $n \leq 4$.

UDGs appear to be quenched systems and occupy the red sequence in clusters
\citep{vandokkum2015a, vdb2016}, although UDG-like systems in the field are typically bluer
\citep[e.g.][]{leisman2017, roman2017b}.
The axial ratios of UDGs in Coma are consistent with being prolate shaped,
with a mean axial ratio of ${\sim}0.7$ and very few with axial ratios less
than 0.4 \citetext{\citealp{burkert2017}, but see \citealp{rong2019}}.
This, along with their low $V_\mathrm{rot}/\sigma$ \citep{vandokkum2019b},
suggests that UDGs are dispersion-dominated systems and not rotationally
supported thick disks, and are perhaps related to the lower mass low surface
brightness dwarf spheroidals in the Local Group \citep{burkert2017}.

It was first proposed that UDGs are ``failed" $L^*$ galaxies after
having lost their gas supply in early times and are now extremely dark matter
dominated, allowing them to survive in such dense environments
\citep{vandokkum2015a}.
Two objects in Coma, Dragonfly 17 and Dragonfly 44, and one in Virgo,
VCC~1287, are possible examples of such a scenario.
Dragonfly~17 hosts ${\sim}30$ globular clusters (GCs), which is abnormally high for its
luminosity, suggesting it could possibly be a ``failed" M33- or LMC-like galaxy
\citep{peng2016, beasley2016b}.
VCC~1287 is similarly a possibly failed LMC-like galaxy based on its GC
system \citep{beasley2016a}.
Dragonfly~44 is much more massive, with ${\sim}75$ GCs and a dynamical mass
consistent with being a failed Milky Way-like system \citep{vandokkum2016,
vandokkum2019b}.
Subsequent investigations, however, suggest that many UDGs have stellar masses and dark
matter halos consistent with dwarf galaxies, suggesting Dragonfly~44 may be an
extreme case \citep{beasley2016a, peng2016, amorisco2018, alabi2018,
ferre-mateu2018}.
At the other extreme, two UDGs in the NGC~1052 group, NGC1052-DF2 and
NGC1052-DF4, have very low velocity dispersions and dynamical masses
consistent with little or no dark matter \citep{vandokkum2018a, vandokkum2019a,
danieli2019a}.

With such a range of properties exhibited by UDGs, the low surface brightness universe is
proving to be just as diverse as the high surface brightness universe.
Numerous formation channels have been brought forward to create such an array of
objects.
In the failed massive halo scenario, possible mechanisms that could remove the
gas and prevent the formation of a normal stellar population include extreme
feedback from supernovae and young stars \citep{agertz2016, dicintio2017,
chan2018}, ram
pressure stripping \citep{yozin2015, jiang2019, tremmel2019} or AGN feedback
\citep{reines2013}.
With these likely being extreme examples however, other mechanisms may also be in
play to form UDGs within lower mass dark matter halos.  If UDGs are ``inflated
dwarfs", both dark matter halos with anonymously high spins
\citep{amorisco2016, rong2017, liao2019} and tidal interactions
\citep{yozin2015, jiang2019, liao2019, martin2019, sales2019} may be responsible for
their large sizes. 
Rare UDGs without dark matter may be the result of high-velocity collisions of
dwarfs in protogroup environments \citep{silk2019}.
Furthermore, the near-linear relation between the abundance of UDGs and
cluster halo mass \citep{vdb2016, roman2017b, vdb2017, janssens2017, mancerapina2018}, in
addition to their existence in low density environments, suggests an
`internal' mechanism of UDG formation that is independent of environment, and
that UDGs are a consistent fraction of the galaxy population in all
environments \citep{amorisco2018}.

At the other extreme of low stellar mass galaxies lie the `ultra-compact
dwarfs' (UCDs).
With characteristic luminosities $\gtrsim 10^7~L_{\odot}$ and radii $r_h
\gtrsim 10~\mathrm{pc}$, they resemble both the nuclei of low-mass galaxies
and the most massive GCs \citep{brodie2011, norris2014, forbes2014, janz2016}.
As they are typically found in the densest environments, environmental
effects, such as tidal stripping, are thought to be involved in their
formation \citep{bekki2003, pfeffer2013}.

In a previous paper, we looked at the UDGs and UCDs in Abell~2744
\citep{janssens2017}, the first cluster observed by the Frontier Fields (FF)
program with the \textit{Hubble Space Telescope} (\textit{HST}).
In this paper, we now investigate the UDGs and UCDs inhabiting all six FF
galaxy clusters.
In addition to being the most massive and distant systems in which UDGs have
yet been discovered, the existing lensing and X-ray analyses permit detailed
study of their local environments.
This paper is organized as follows.
In Section \ref{sec:data}, we describe the FF program and its data.
Our methods, including UDG and UCD selection, are described in Section
\ref{sec:methodology}.
In Section \ref{sec:results}, we present and discuss the results of our
analysis, primarily the abundance of UDGs in the six FF clusters and their
spatial distributions in relation to other classes of galaxies and known
substructures in the clusters.

We adopt a $\Lambda$CDM cosmology with $\Omega_m = 0.3$, $\Omega_{\Lambda} =
0.7$, $H_0 = 70~\mathrm{km}~\mathrm{s}^{-1}~\mathrm{Mpc}^{-1}$. All magnitudes
are in the AB system. Galactic extinction corrections from the
\cite{schlafly2011} extinction maps were applied to all colours and
magnitudes.\footnote{Using the online calculator at
\url{https://ned.ipac.caltech.edu/forms/calculator.html}.}

\section{Data}\label{sec:data}

The \textit{HST} FF program has produced the deepest images to date of galaxy
clusters and gravitationally lensed galaxies for six high-magnification
clusters---
Abell~2744, MACSJ0416.1$-$2403, MACSJ0717.5$+$3745, MACSJ1149.5$+$2223,
Abell~S1063 (also known as RXCJ2248.7$-$4431) and Abell~370---
along with six corresponding parallel ``blank" fields offset
${\sim}6\arcmin$ from each cluster \citep{lotz2017}.
These clusters were chosen for their known high lensing strengths, low sky
backgrounds and Galactic extinctions, in addition to observability with
\textit{HST}, \textit{Spitzer} and ground-based facilities \citep{lotz2017}.
Selecting galaxy clusters for their lensing strength will end up selecting
extremely massive, merging clusters, as the merger stretches the lensing
critical curves between the various components resulting in relatively large
areas subject to high magnification \citep{redlich2012, diego2016}.
The cluster properties are summarized in Table \ref{tab:clusters}.
The coordinates are the cluster centres as defined by their stellar content,
that is the location of the brightest cluster galaxy (BCG), or where the
cluster is comprised of multiple merging subclusters, the centroid of the BCGs.
The diverse and complex morphologies displayed by these clusters is discussed
later.
Each cluster and parallel field pair was observed for 70 orbits with the
Advanced Camera for Surveys Wide Field Camera (ACS/WFC) in $F435W$, $F606W$
and $F814W$, and 70 orbits with the Wide Field Camera 3 IR channel (WFC3/IR)
in $F105W$, $F125W$, $F140W$ and $F160W$, achieving $5\sigma$ depths of
${\sim}29$th AB magnitude \citep{lotz2017}.

Despite the ${\sim}6\arcmin$ separations between the cluster and parallel fields,
the parallel fields are still either within or straddle the virial radii
($R_{200}$) of the clusters and so these images are examined for UDGs as well.
To estimate the contamination of our UDG sample by background galaxies, we
instead use the eXtreme Deep Field \citep[XDF,][]{illingworth2013}.
This is the only image of the sky that is deeper than the FFs to date, and
was obtained by stacking the data from 19 different \textit{HST} programs
completed between 2002 and 2012 covering the \textit{Hubble} Ultra Deep Field
\citep{illingworth2013}.
The XDF has ACS/WFC coverage in $F435W$, $F606W$, $F775W$, $F814W$ and $F850LP$,
and WFC3/IR coverage in $F105W$, $F125W$, $F140W$ and $F160W$.

\begin{deluxetable*}{lrrrrrrr}
\tablecaption{Frontier Fields Cluster Properties and UDG Abundances}
\tablecolumns{8}
\tablewidth{0pc}
\tablehead{
    \multicolumn{1}{l}{Cluster} &
    \multicolumn{1}{r}{$z_\mathrm{cl}$} &
    \multicolumn{1}{r}{R.A.\tablenotemark{a}} &
    \multicolumn{1}{r}{Dec.\tablenotemark{a}} &
    \multicolumn{1}{c}{$M_{200}$\tablenotemark{b}} &
    \multicolumn{1}{c}{$R_{200}$\tablenotemark{b}} &
    \multicolumn{2}{c}{Number of UDGs}
    \\
    \colhead{} &
    \colhead{} &
    \multicolumn{1}{r}{(J$2000.0$)} &
    \multicolumn{1}{r}{(J$2000.0$)} &
    \multicolumn{1}{c}{($10^{15}~M_\odot$)} &
    \multicolumn{1}{c}{(Mpc)} &
    \multicolumn{1}{r}{Raw\tablenotemark{c}} &
    \multicolumn{1}{r}{Total\tablenotemark{d}}
}
\startdata
    Abell 2744                   & $0.308$ & 00:14:20.70 & $-$30:24:00.58 & $2.06 \pm 0.42$ & $2.35 \pm 0.16$ & 99  & $1351^{+387}_{-379}$ \\
    Abell S1063\tablenotemark{e} & $0.348$ & 22:48:43.97 & $-$44:31:51.14 & $1.88 \pm 0.67$ & $2.38 \pm 1.48$ & 167 & $1416^{+1877}_{-1127}$ \\
    Abell 370                    & $0.375$ & 02:39:52.94 & $-$01:34:37.00 & $3.16 \pm 0.38$ & $2.66 \pm 0.11$ & 65  & $711^{+213}_{-210}$ \\
    MACSJ0416.1$-$2403           & $0.396$ & 04:16:08.38 & $-$24:04:20.80 & $1.07 \pm 0.26$ & $1.88 \pm 0.69$ & 66  & $219^{+230}_{-164}$ \\
    MACSJ1149.5$+$2223           & $0.543$ & 11:49:35.70 & $+$22:23:54.73 & $2.50 \pm 0.54$ & $2.35 \pm 1.00$ & 109 & $582^{+397}_{-364}$ \\
    MACSJ0717.5$+$3745           & $0.545$ & 07:17:32.63 & $+$37:44:59.70 & $2.68 \pm 0.55$ & $2.36 \pm 0.77$ & 91  & $609^{+438}_{-359}$
\enddata
\tablenotetext{a}{Cluster centres were adopted as follows:
    A2744, location of BCG nearest X-ray centroid;
    AS1063, location of BCG;
    A370, midpoint between BCGs \citep{lagattuta2017};
    M0416, midpoint between BCGs \citep{zitrin2013};
    M1149, location of BCG;
    and
    M0717, mean location of red sequence members \citep{medezinski2013}.
    }
\tablenotetext{b}{Determined from gravitational lensing analyses. $M_{200}$
    and $R_{200}$ for A2744 are from \cite{medezinski2016}, A370 from
    \cite{umetsu2011}. Values for the other clusters are from
    \cite{umetsu2016}.}
\tablenotetext{c}{Total number of UDGs detected in the cluster and parallel
    fields.}
\tablenotetext{d}{Estimate of the total number of UDGs within $R_{200}$ after
    background, completeness and geometrical corrections.}
\tablenotetext{e}{Also known as RXCJ2248.7$-$4431.}
\label{tab:clusters}
\end{deluxetable*}

\section{Methodology}\label{sec:methodology}

\subsection{Source detection}

For each of the six FF clusters, \textsc{SExtractor} \citep{bertin1996} was run
in dual image mode on the 30 mas images of both the cluster core and parallel
fields. The $F814W$ image was used as the detection image for all bands.
A \texttt{WEIGHT\_THRESH} of 0.002 was used to remove sources detected in the
low exposure time regions along the edges of the fields.
The XDF was treated similarly.
In \cite{janssens2017}, we used the $F775W$ image as the XDF detection image
since it is much deeper than the XDF $F814W$ image. However, the XDF $F814W$ image
is of comparable depth to the FF $F814W$ images, and its use simplifies the
background correction while having no effect on the results.
Only 60 mas WFC3/IR images are available for the XDF so a second 60 mas multiband
catalog was created for the XDF, matched to the 30 mas catalog, and used only
for WFC3/IR colours.

\subsection{Structural parameters}\label{sec:galfit}

GALFIT \citep{peng2002} was used to determine $I_{814}$-band structural parameters
by fitting a single component S\'{e}rsic model to the $F814W$ image of every
object brighter than $F814W = 28$ mag with \textsc{SExtractor} $\mathtt{FLAGS}
< 4$ in the catalogs.

We used \textsc{PSFEx} \citep{bertin2011} to supply GALFIT with point spread
functions (PSFs) for each object. For each image, a sample of point sources
was selected from a shallower \textsc{SExtractor} catalog with the cuts $1.0 <
\mathtt{FWHM} < 10.0$ pixels, signal-to-noise ratio $\geq 5$ and ellipticity $e < 0.3$.
A cubic polynomial was used to map the variability of the PSF across the
image.

Similar to \cite{vanderwel2012}, the provided reduced images and RMS maps were
combined to produce total noise maps to pass to GALFIT. The RMS maps created
by Drizzle account for the ``instrinsic" sources of noise, e.g.\ dark current,
readout noise, and background noise \citep{koekemoer2011}. The Poisson noise
from the sources themselves are readily computed from the images, and are
added to the RMS maps in quadrature.  Where available, the exposure time maps
were used to convert the images, in electrons per second, to electrons for
this computation. Otherwise, the median exposure time was used instead.

A ``segmentation vector" was then created which, for each pixel, lists the IDs
of objects which contribute light to that pixel. This was done by stacking the
Kron ellipses (semi-major axis equal to $2 \times \mathtt{A\_IMAGE} \times
\mathtt{KRON\_RADIUS}$ from \textsc{SExtractor}) of every object in the image.
Ellipses are readily generated from the \texttt{CXX\_IMAGE},
\texttt{CYY\_IMAGE} and \texttt{CXY\_IMAGE} ellipse parameters. Neighbours
which overlap with a given object are then the unique set of IDs within this
object's Kron ellipse.  For each object, a cutout large enough to contain all
overlapping neighbours was created from both the image and the total noise
map.  Overlapping neighbours are fit simultaneously with the object in
question.  Bad pixels and non-overlapping neighbours were masked by supplying
GALFIT with a mask image where bad pixels and pixels within the Kron ellipses
of irrelevant sources were given a value of 1, with all other pixels 0.
Finally, a PSF image was created at the object's position from the \textsc{PSFEx}
model.

Absolute magnitudes and physical sizes were computed from the GALFIT model
parameters assuming all sources in an image lie at their respective cluster
redshifts (Table \ref{tab:clusters}).  We circularized the effective radii
using $R_{e,c} = R_e\sqrt{b/a}$. The mean surface brightness within $R_e$ was
derived from the model magnitudes $m$ and effective radii using 
\begin{equation}
    \langle\mu\rangle_{e} = m + 2.5\log(2 \pi R_{e,c}^2).
\end{equation}
The absolute mean surface brightness is then calculated with
\begin{equation}
    \langle\mu\rangle_{e,abs} = \langle\mu\rangle_{e} -
    2.5\log(1+z)^4 - E(z) - K(z),
\end{equation}
where $z$ is the cluster redshift, and $E(z)$ and $K(z)$ are the evolutionary
and $K$-corrections, respectively \citep{graham2005}, computed with EzGal
\citep{ezgal} assuming a simple stellar population (SSP) with
$[\mathrm{Fe}/\mathrm{H}] = -0.6$, a formation redshift of $z=2$ and a
\cite{chabrier2003} initial mass function.

\subsection{Ultra-diffuse galaxy selection}\label{sec:udgselection}

\begin{figure*}
	\includegraphics[width=0.99\textwidth]{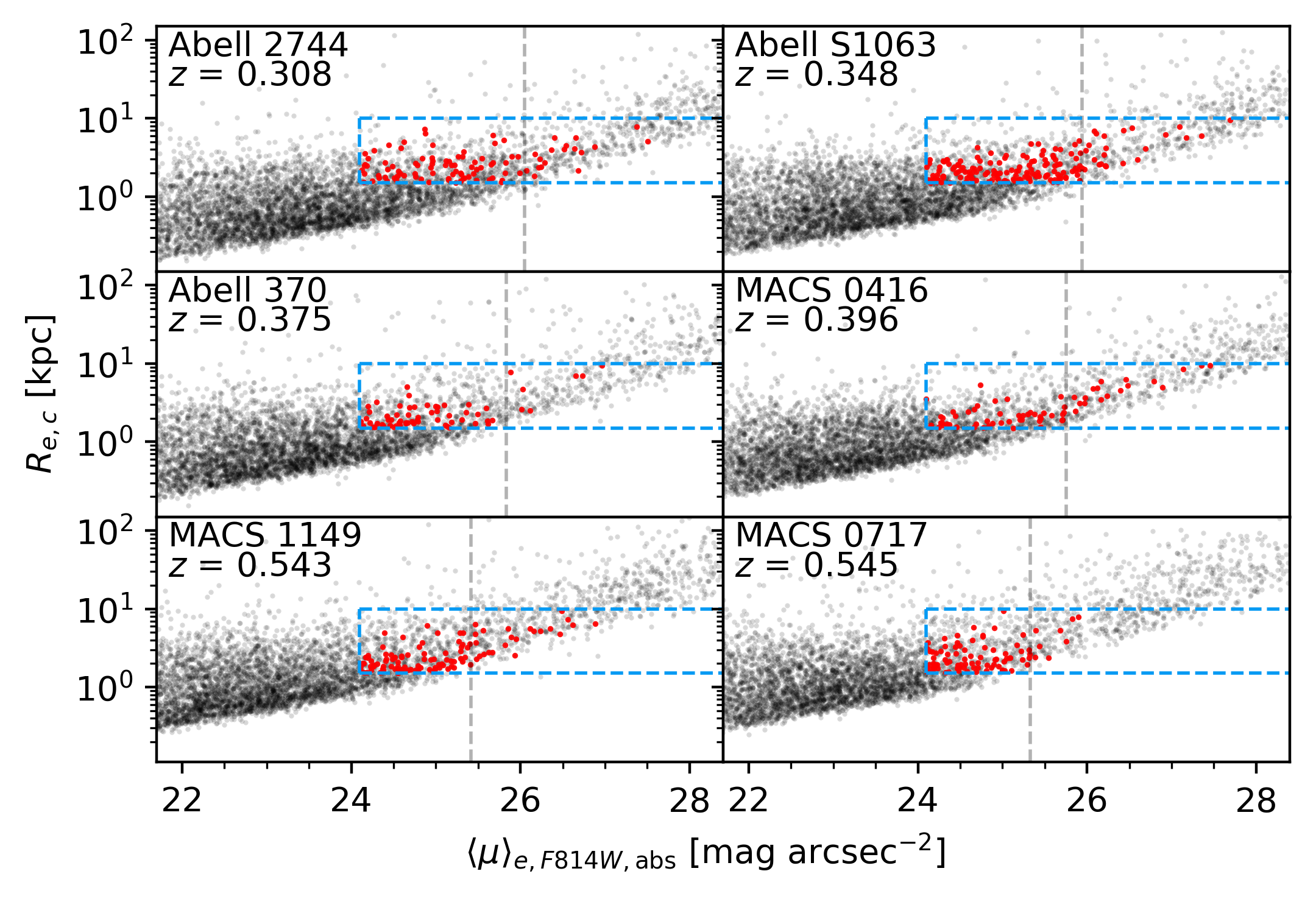}
	\caption{
    UDG selection by size and absolute mean surface brightness within $R_e$.
    Effective radii and absolute surface brightnesses are computed assuming
    all detected sources reside at the cluster redshift. The UDGs, shown in
    red, are selected with $\langle\mu\rangle_{e,\mathrm{abs},F814W} > 24.1 ~
    \mathrm{mag}~\mathrm{arcsec}^{-2}$ and circularized effective radii in the
    range $1.5~\mathrm{kpc} \leq R_{e,c} < 10~\mathrm{kpc}$ (blue dashed lines). The additional
    cuts, including a visual inspection,
    are described in the text. The dashed vertical grey line corresponds to
    the 50\% mean surface brightness completeness limit of
    $26.9~\mathrm{mag}~\mathrm{arcsec}^{-2}$, transformed to the redshift of
    each cluster.
    \label{fig:udg_selection}
	}
\end{figure*}

Our UDG selection is very similar to that used in our previous work on
Abell~2744 \citep{janssens2017}, which in turn is based on the cuts used by
\cite{vdb2016}.
In this work, instead of transforming all the UDG $F814W$ surface brightnesses
to the $r$-band, we instead transform the $r$-band surface brightness cut of
$\langle\mu\rangle_{e,abs,r} > 23.8~\mathrm{mag}~\mathrm{arcsec}^{-2}$ into
a $F814W$ cut using the same SSP described above for the evolutionary and
$K$-corrections.
The cuts are as follows:
\begin{itemize}
\item Circularized half-light radius in the range $1.5~\mathrm{kpc} \leq R_{e,c} <
    10~\mathrm{kpc}$.
\item S\'{e}rsic index $n < 4$. Roughly 97\% of injected $n=1$ profiles have a
    recovered $n < 4$ (see \S \ref{sec:sims}).
\item Absolute mean surface brightness within $R_e$
    $\langle\mu\rangle_{e,\mathrm{abs},F814W} >
        24.1~\mathrm{mag}~\mathrm{arcsec}^{-2}$. For radial profiles and
        estimating the total abundance in each cluster, only UDGs with a mean
        surface brightness within $R_e$ brighter than
        $\langle\mu\rangle_{e,F814W} = 26.9~\mathrm{mag}~\mathrm{arcsec}^{-2}$
        are included, this is the 50\% completeness limit (see \S
        \ref{sec:sims}).
\item Axis ratio $q \equiv b/a > 0.3$. Coma UDGs are prolate with $\langle q
    \rangle \sim 0.7$ \citep{burkert2017} and \cite{chen2010} found no Virgo
        dwarfs flatter than 0.35. This also has the benefit of removing
        edge-on disks and lensing arcs.
\item Photometric redshift $z_{\mathrm{phot}} < 1$, if available. Not every
    UDG candidate has a match in the ASTRODEEP catalogs, but this removes
    known high-$z$ background objects.
\item Within the WFC3/IR footprint for uniform photometric redshift accuracy.
\end{itemize}
The selection in size and surface brightness parameter space is shown in Figure
\ref{fig:udg_selection}.

\begin{figure}
	\includegraphics[width=0.50\textwidth]{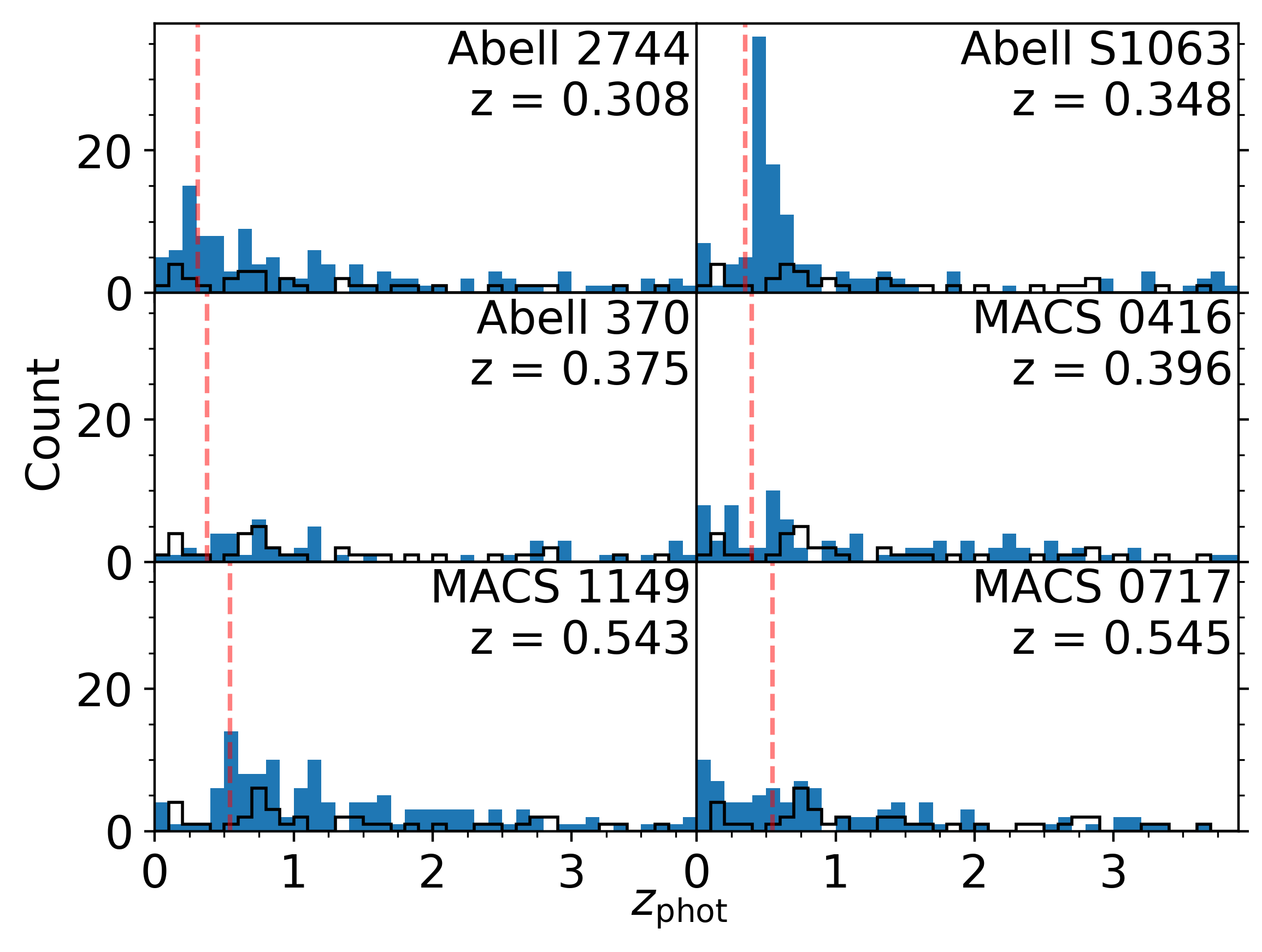}
	\caption{
    The \textsc{ASTRODEEP} photo-$z$ distributions of sources that pass our
    UDG selection cuts are shown in blue (sources with $z_\mathrm{phot} \geq
    1$ were not visually inspected).
    Objects in the XDF that pass the cuts for each cluster are shown in black.
    The vertical line is the redshift of the cluster.
    \label{fig:photo-z}
	}
\end{figure}

Photometric redshifts for all six clusters were obtained from the
\textsc{ASTRODEEP} catalogs \citep{castellano2016,merlin2016,bradac2019}.
We matched our catalog to that of \textsc{ASTRODEEP} by finding the nearest
neighbour within 0.15\arcsec, or 5 ACS pixels.
The distribution of photo-$z$'s of sources that pass the UDG selection cuts
is shown in Figure \ref{fig:photo-z}.
Photo-$z$'s were not used in the computation of physical sizes and absolute
magnitudes since not every UDG has a match in the \textsc{ASTRODEEP} catalog.

\begin{figure*}
	\includegraphics[width=0.50\textwidth]{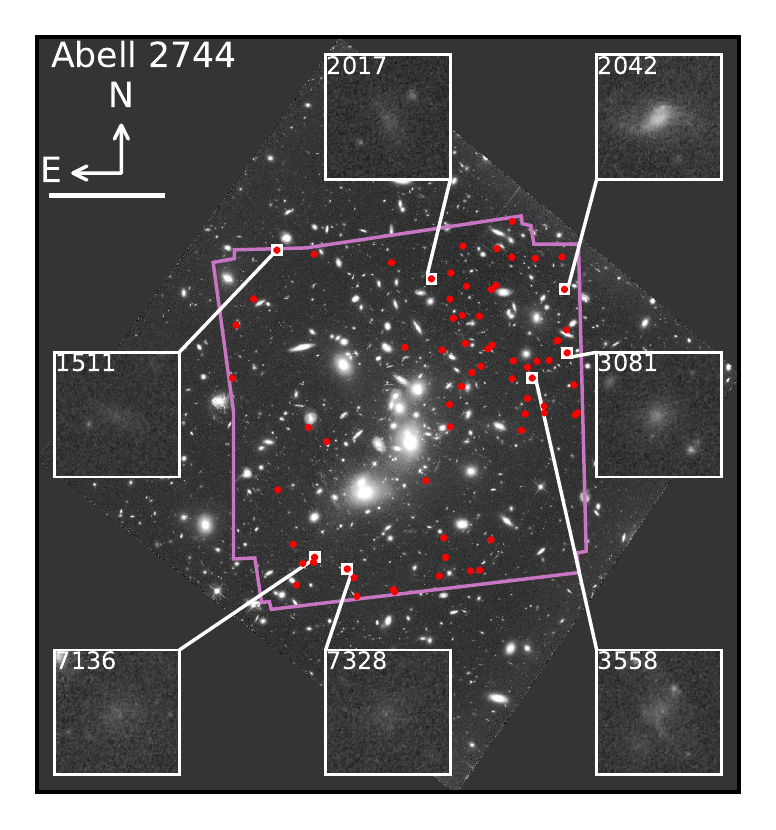}
    \includegraphics[width=0.50\textwidth]{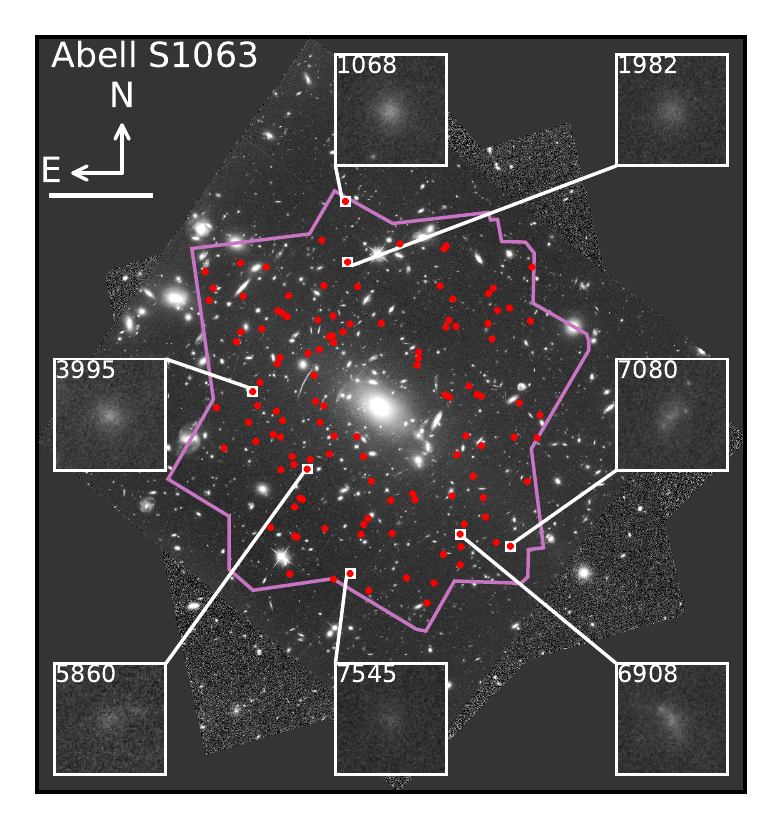}
	\caption{
    Locations of selected UDGs (red points) in the Abell~2744 and Abell~S1063
    cluster core fields.
    The background image is the ACS $F814W$ image and the pink border is the
    extent of the WFC3/IR coverage.
    North is up and east is to the left.
    The bar below the compass corresponds to 200 kpc. 
    Insets show zoom-ins on select UDGs, the sizes of which are 15 kpc a side.
    \label{fig:locations}
	}
\end{figure*}
\begin{figure*}
    \ContinuedFloat
	\includegraphics[width=0.50\textwidth]{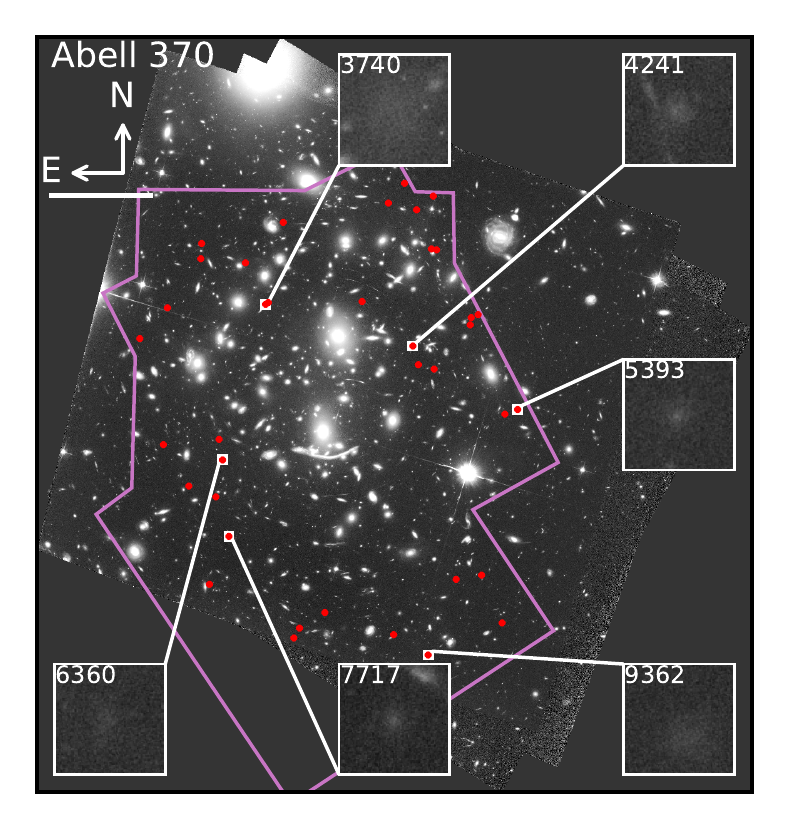}
	\includegraphics[width=0.50\textwidth]{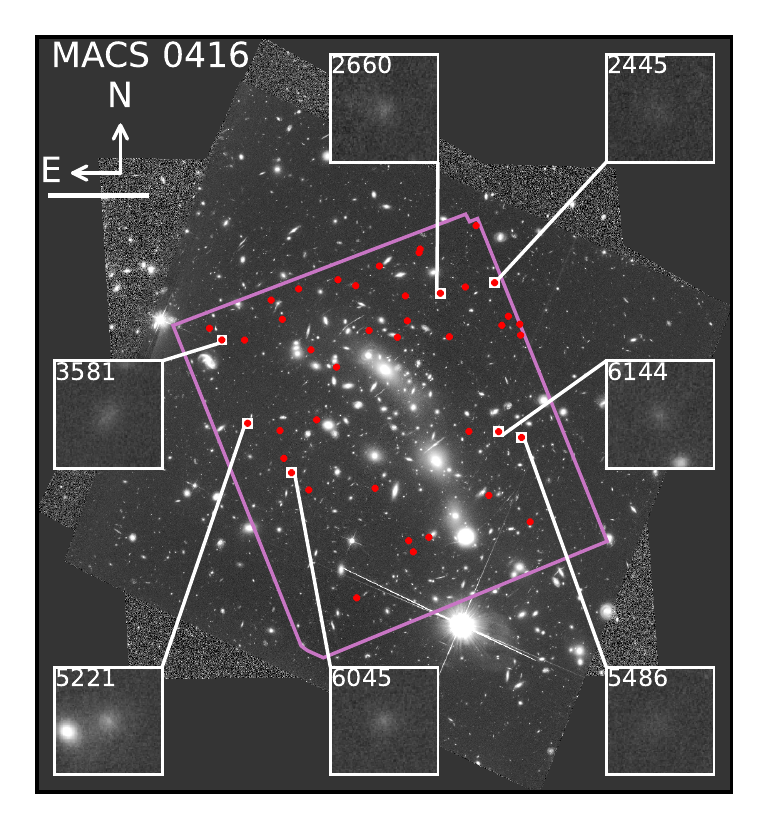}
    \caption{(Continued.) Abell~370 and MACS~0416 cluster core fields.}
\end{figure*}
\begin{figure*}
    \ContinuedFloat
	\includegraphics[width=0.50\textwidth]{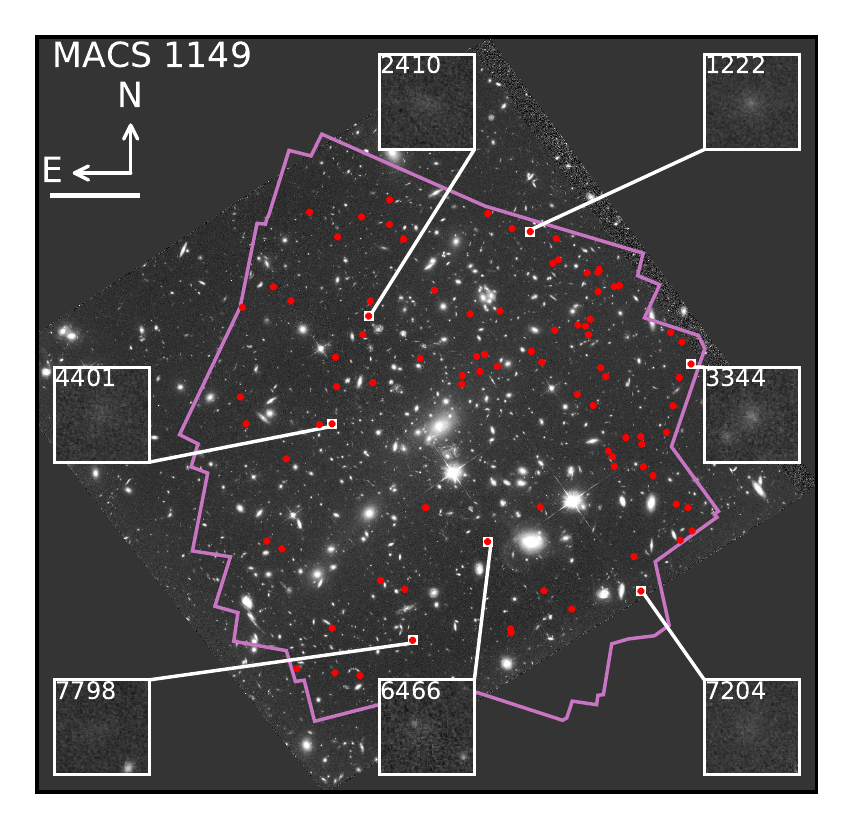}
	\includegraphics[width=0.50\textwidth]{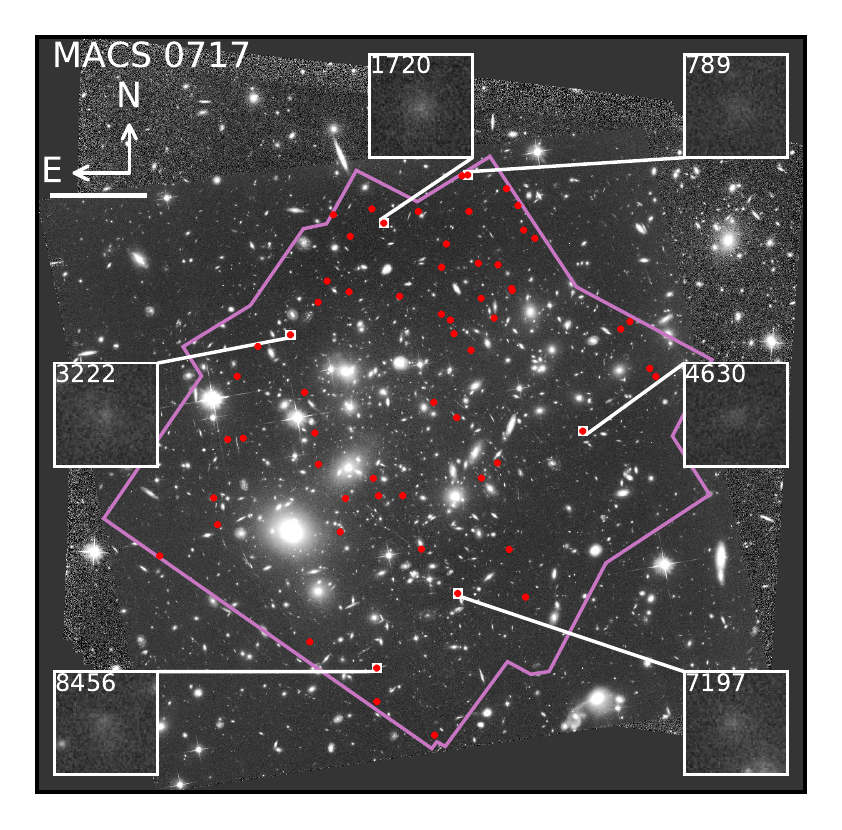}
    \caption{(Continued.) MACS~1149 and MACS~0717 cluster core fields.}
\end{figure*}

All UDG candidates were visually inspected. Of the 1190 candidates, 636
UDGs were kept. The most common contaminants are deblended objects (i.e.\
spiral arms, lensing arcs and tidal features split into multiple objects by
\textsc{SExtractor}) and compact ($n \sim 3.5$) galaxies with high central
surface brightnesses that do not visually resemble UDGs. Finally, 18 duplicate
UDGs were removed where a UDG was deblended into multiple sources, in which
case the brighter object was kept.
In a few of these cases, \textsc{SExtractor} picked out an offset overdensity,
similar in appearance to DGSAT~I \citep{martinez-delgado2016}.
The raw number of UDGs detected in each cluster is listed in Table
\ref{tab:clusters}.
Figure \ref{fig:locations} shows the locations of all selected UDGs within the
WFC3 coverage (pink outline) for each cluster core field, along with $15
\times 15$ kpc zoom-ins on select UDGs.

Note the remarkable non-uniform projected spatial distributions of UDGs in some
of the clusters, most notably in Abell~2744, MACS~1149 and MACS~0717.
Only in Abell~S1063 and Abell~370 do the UDGs appear to be evenly distributed
around the cluster. In the other clusters, there appear to be many more UDGs
on one side than the other.
UDGs also appear to avoid the central regions of the clusters.
For now though, we turn to corrections and simulations needed to understand
the physical significance of these effects, if any.

\subsection{Background correction}

For each FF cluster, an estimate of the background contamination of the UDG
sample is made by computing the physical sizes and absolute surface
brightness of every XDF source assuming the source lies at the FF cluster
redshift.
We then apply the UDG cuts described above.
By assuming that all XDF sources lie at the FF cluster redshift of $z =
0.308$--$0.545$, the assumed physical scale together with $(1 + z)^4$
cosmological dimming conspire to turn distant high-$z$ galaxies into objects
consistent with UDGs.
In order to remove these from the background correction count, our XDF catalog
was matched to the UVUDF photo-$z$ catalog of \cite{rafelski2015}.
The photo-$z$ distributions of XDF sources that pass the UDG cuts in each
cluster are shown as the black histograms in Figure \ref{fig:photo-z}. 

\subsection{Image simulations}\label{sec:sims}

\begin{figure}[t]
	\includegraphics[width=0.50\textwidth]{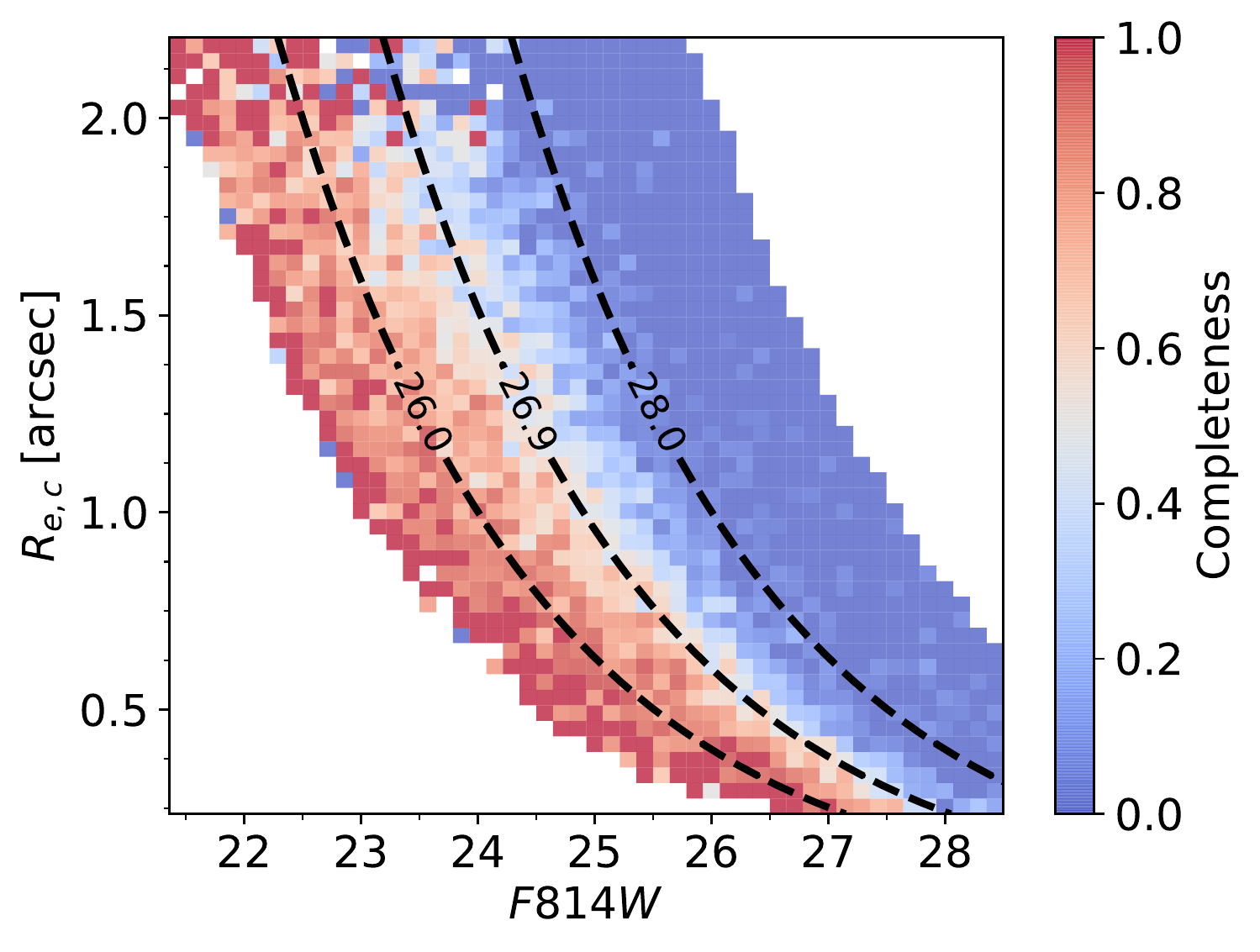}
	\caption{
    \label{fig:completeness}
    Recovery fraction as a function of size and magnitude for $n=1$ S\'{e}rsic
    profiles. The lines are contours of mean surface brightness within $R_e$
    ($\langle\mu\rangle_{e,F814W}$) in units of
    $\mathrm{mag}~\mathrm{arcsec}^{-2}$; 26.9
    $\mathrm{mag}~\mathrm{arcsec}^{-2}$ is the 50\% completeness limit.
	}
\end{figure}

Two sets of image simulations were performed. The first was to determine and
inform the selection criteria used to select UDGs, and the second was to determine
geometrical completeness distributions of objects that pass our selection
criteria in the six clusters.

We began by injecting 2000 simulated objects into each of the cluster and
parallel image pairs at random positions in the region with overlapping ACS
and WFC3 coverage. 
The \textsc{SExtractor} segmentation map was used to ensure that a chosen location
was empty.
To prevent crowding, this was done in batches of 20, with no pair of injected
objects permitted to be closer than 150 pixels.
To ensure enough objects were injected at small radii, positions were
chosen in radial coordinates for the cluster fields.
The simulated objects were single S\'{e}rsic profiles generated using
\textsc{GALFIT} with the following parameters: S\'{e}rsic index $n = 1$,
circularized effective radius $1.5~\mathrm{kpc} \leq R_{e,c} < 10~\mathrm{kpc}$, central surface brightness
$17~\mathrm{mag}~\mathrm{arcsec}^{-2} < \mu_0 < 29~\mathrm{mag}~\mathrm{arcsec}^{-2}$, axis ratio $0.3 \leq
b/a < 1.0$, and position angle $0^{\circ} \leq \theta < 360^{\circ}$.  Sets of parameters that
resulted in objects much too bright to be a UDG
($\langle\mu\rangle_{e,\mathrm{abs},F814W} <
23.5~\mathrm{mag}~\mathrm{arcsec}^{-2}$) or far too faint to be reliably
detected ($m_{F814W} > 28.5$ or $\langle\mu\rangle_{e,F814W} >
29.5~\mathrm{mag}~\mathrm{arcsec}^{-2}$) were thrown out and redrawn.
The simulated images were then analyzed with the same pipeline described
above, with the exception that only the nearest detected object within 5
pixels of an injected location was selected for \textsc{GALFIT} fitting. Roughly 97\%
of recovered objects have a S\'{e}rsic index $n < 4$.
Figure \ref{fig:completeness} shows the recovery fraction as a function of
effective radius and magnitude. We find a 50\% completeness limit of
$\langle\mu\rangle_{e,F814W} = 26.9~\mathrm{mag}~\mathrm{arcsec}^{-2}$ with no
significant variation between the six clusters.

The geometrical completeness simulations are similar, but we only inject objects
that would pass our selection criteria and that are brighter than the 50\%
completeness limit found above. We allow any location in the
ACS and WFC3 overlap region to be chosen, informing the fraction of UDGs that may
be lost due to projection against other sources or the intracluster light
(ICL). The minimum spacing of 150
pixels between injected objects is still enforced. The resulting radial
completeness curves are shown in the right-hand panels of Figure
\ref{fig:radial-profiles}, the rest of the figure is discussed later in the
context of the radial density profiles of UDGs.

\subsection{Ultra-compact dwarf selection}

\begin{figure*}
	\includegraphics[width=1.00\textwidth]{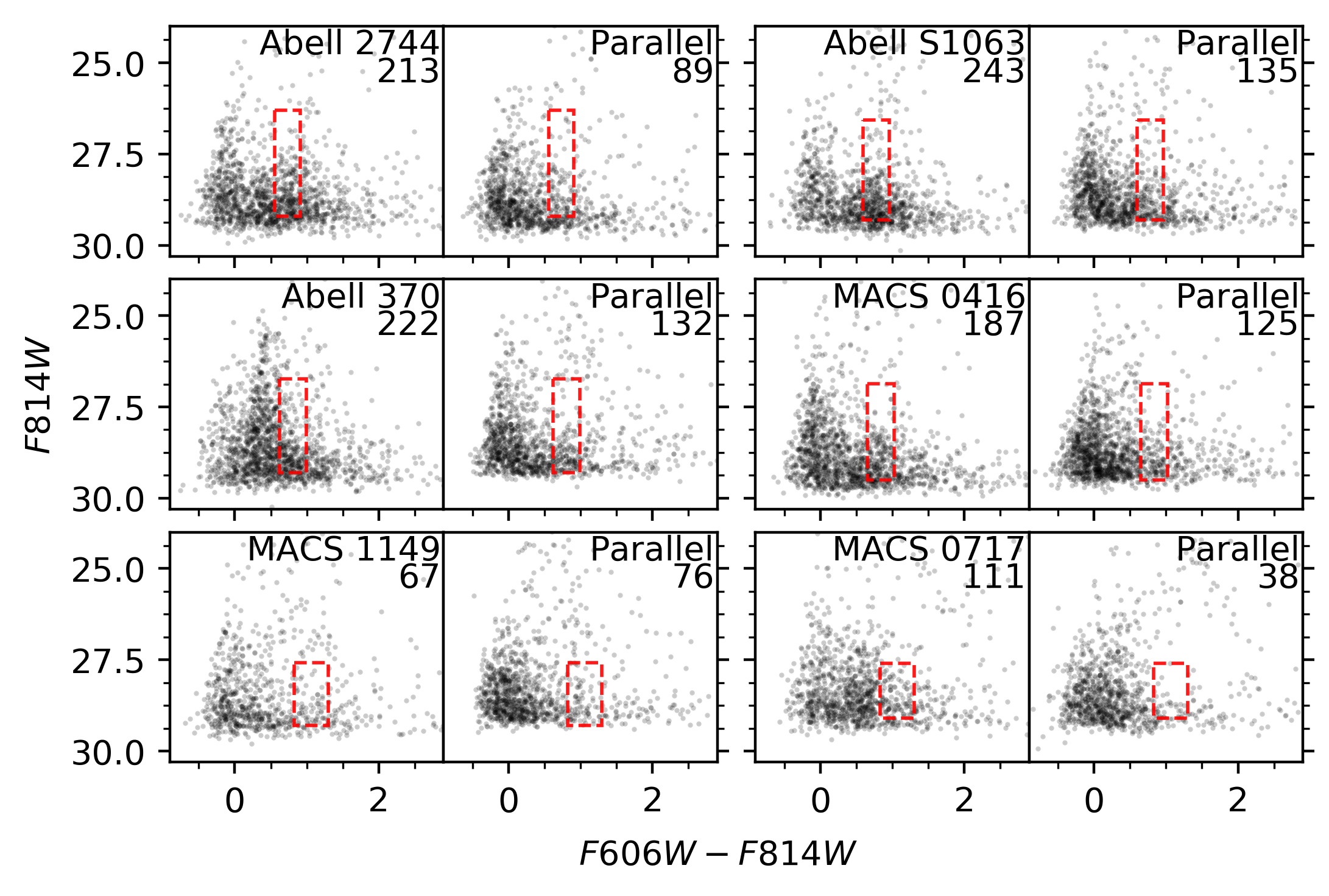}
	\caption{
    \label{fig:cmd}
    Colour-magnitude diagrams for unresolved sources in each of the cluster
    (left panels) and parallel (right) fields.
    The boxes are the UCD selection regions, described in the text. 
    The numbers in the top right are the total number of UCD candidates in
    each field.
    UCDs are expected to be found in the densest regions so the parallel
    fields are used to estimate the contamination from unresolved galaxies and
    foreground Milky Way stars.
	}
\end{figure*}

UCDs were selected from a separate catalog optimized for point source
detection since even the largest UCDs \citetext{$r_h \sim 100~\mathrm{pc}$,
\citealp{brodie2011}} would be
unresolved in the lowest redshift FF cluster.
A median filter with a kernel size of 15 pixels was applied to the $F814W$
image of each field and subtracted off to remove low frequency power from the
BCGs and the ICL.
The resulting image was then used by \textsc{SExtractor} as the detection
image in dual-image mode for all bands, with measurements performed on the
unfiltered images.
Magnitudes were measured in 4 pixel diameter apertures.
Aperture corrections were applied by first correcting to a 1{\arcsec} diameter
(33.3 pixels) aperture and then Table 5 in \cite{sirianni2005} was used to
correct from the 1{\arcsec} aperture to infinity.
Since the PSF varies spatially across each image, the \textsc{PSFEx} model was
used to compute the correction to a 1{\arcsec} aperture at the location of
every detected source.

Point sources were selected from this catalog on the basis of shape,
requiring $e \equiv 1-b/a < 0.4$, size, requiring $\mathtt{FLUX\_RADIUS} <
10~\mathrm{pixels}$, and image concentration, requiring $0.8 < C_{3-7} <
1.2$, where $C_{3-7}$ is the difference between an object's magnitude measured
in a 3 and 7 pixel diameter aperture.
The $F606W,F814W$ colour-magnitude diagrams (CMDs) for point sources in the
FFs are shown in Figure \ref{fig:cmd}.
Since UCDs are expected to be found in the densest environments
\citep{pfeffer2013}, we select UCDs in only the cluster fields and use the
parallel fields to estimate the contamination from unresolved galaxies and
foreground stars.
The boxes shown in Figure \ref{fig:cmd} are the UCD selection regions which
are the apparent $F814W$ magnitudes and $F606W-F814W$ colours spanned by SSPs
with formation redshifts $2 < z < 10$ and metallicities $-2.25 <
[\mathrm{Fe}/\mathrm{H}] < -0.33$ at the redshift of each cluster.
The bright magnitude limit corresponds to a mass of $10^7~M_{\odot}$ and
the faint limit is the 50\% completeness limiting magnitude for the cluster field
determined using artificial star tests (see Appendix \ref{sec:artstartests}).
The limiting magnitudes are well above the magnitudes of the most
massive GCs, with GCs expected to have apparent magnitudes $m_{814} \gtrsim
31$ mag. 
The number of sources that reside within the selection box is listed for each
field in Figure \ref{fig:cmd}.
An excess of UCD candidates in the cluster core field is observed for all
clusters, with the exception of MACS~1149.

In the Abell~370 cluster core field CMD, the sequence of sources bluer than the
UCD selection box at $F606W-F814W \sim 0.5$ are likely GCs associated with the
foreground elliptical galaxy PGC~175370 at a distance of ${\sim}200$ Mpc.
This is the galaxy on the northern edge of the ACS field in Figure
\ref{fig:locations}.
Photometric scatter at the faint end is a source of foreground contamination
not captured by the parallel field.
Restricting the UCD analysis to the region with WFC3 coverage removes the most
likely contaminants (see Figure \ref{fig:pgc175370}).

\section{Results and Discussion}\label{sec:results}

\subsection{Radial distributions}

\begin{figure*}
    \epsscale{0.7}
	\plotone{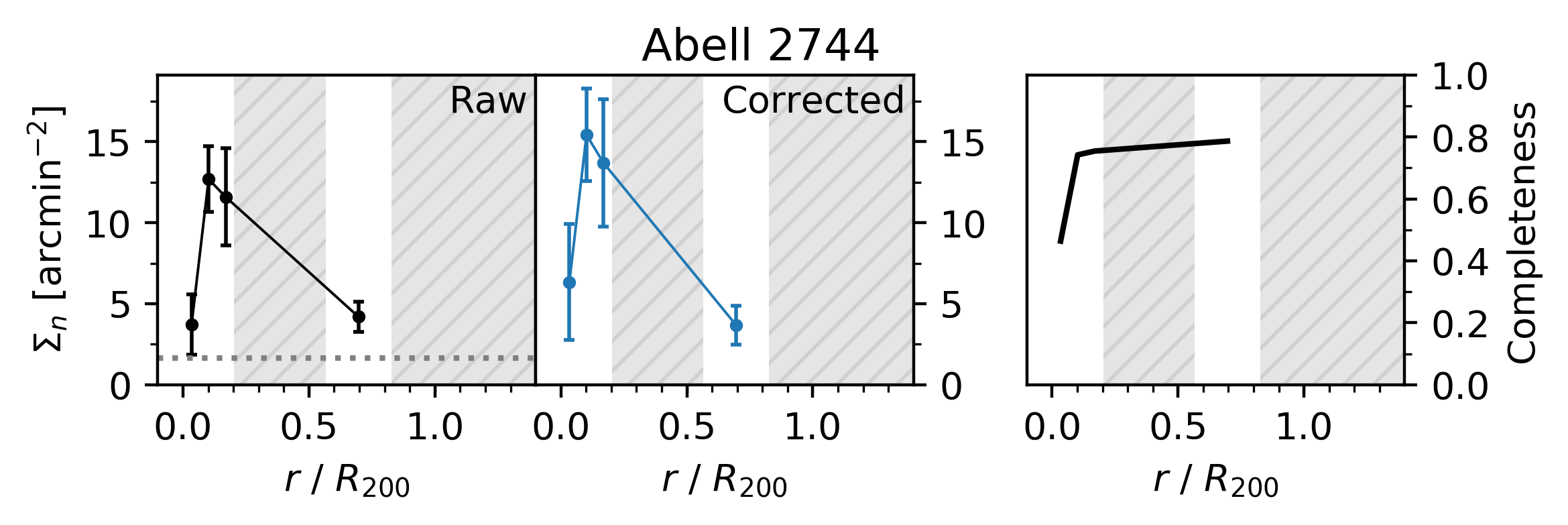}
	\plotone{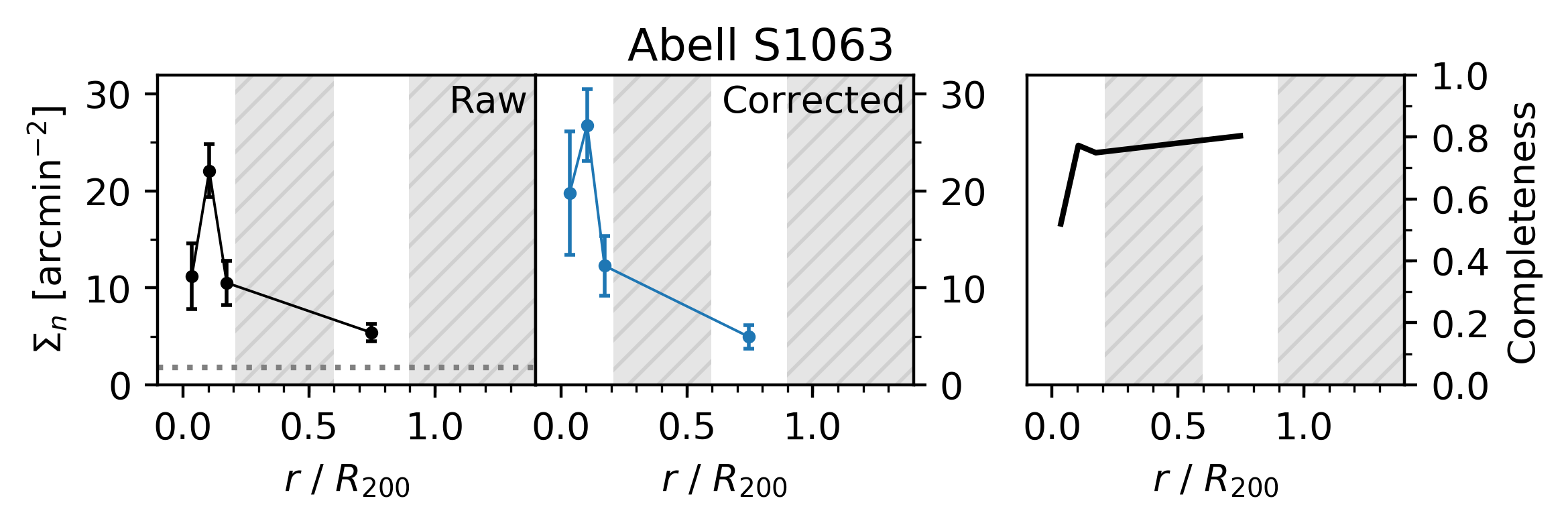}
	\plotone{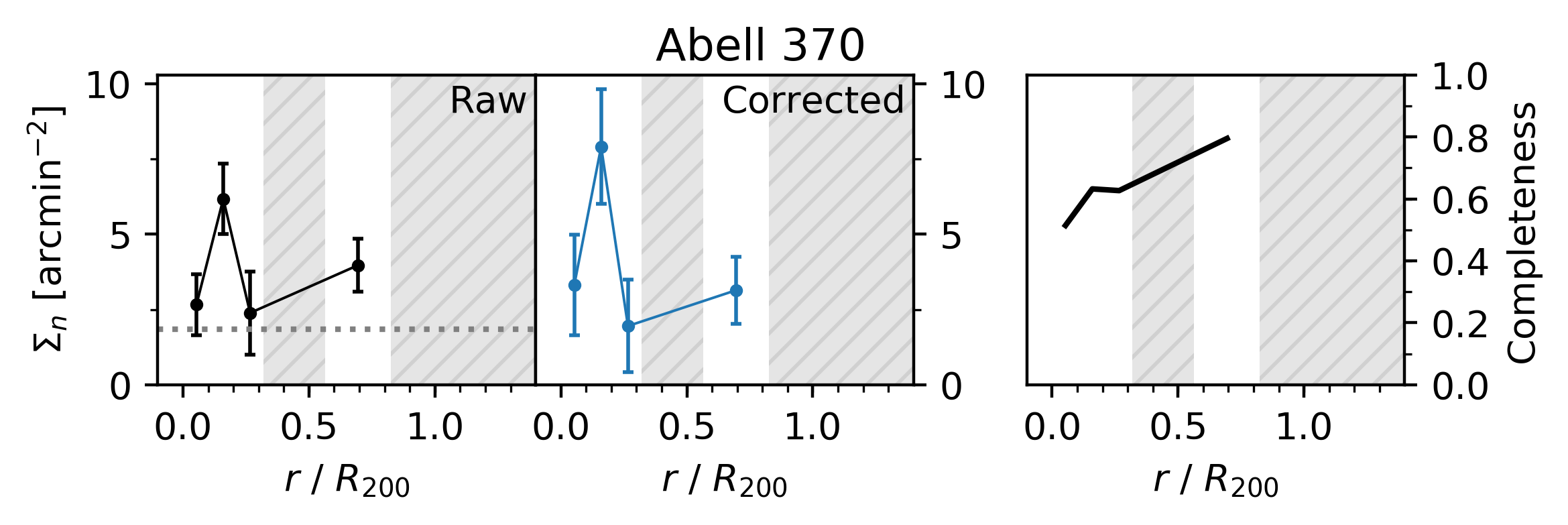}
	\plotone{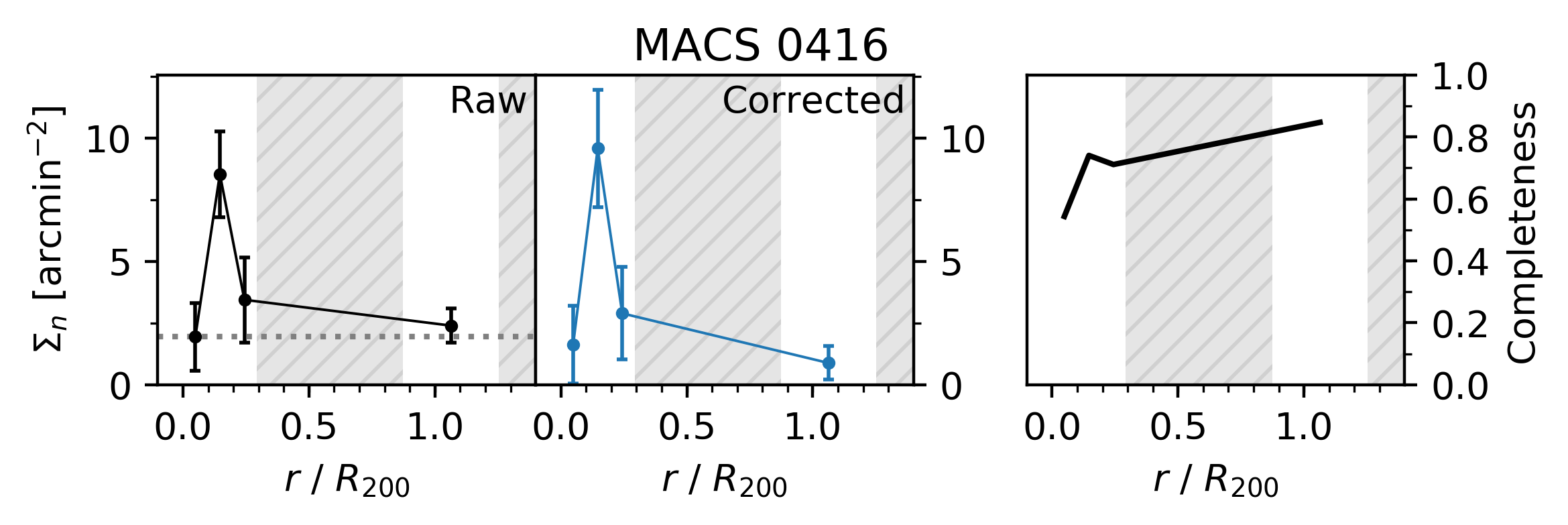}
	\plotone{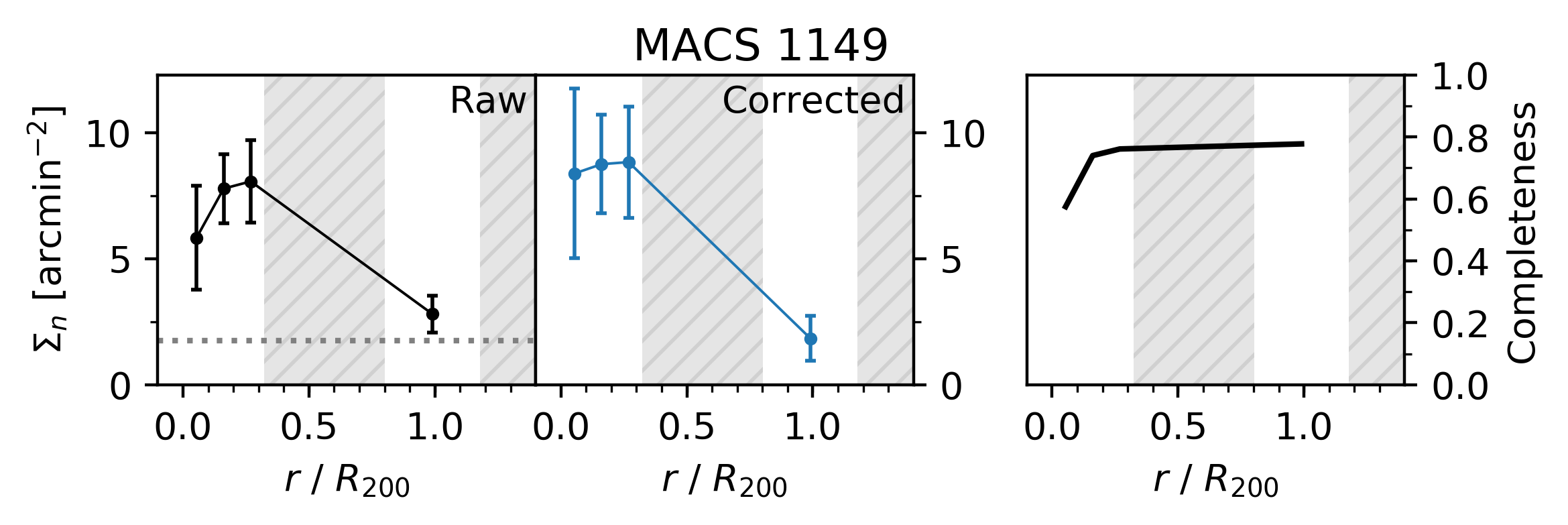}
	\plotone{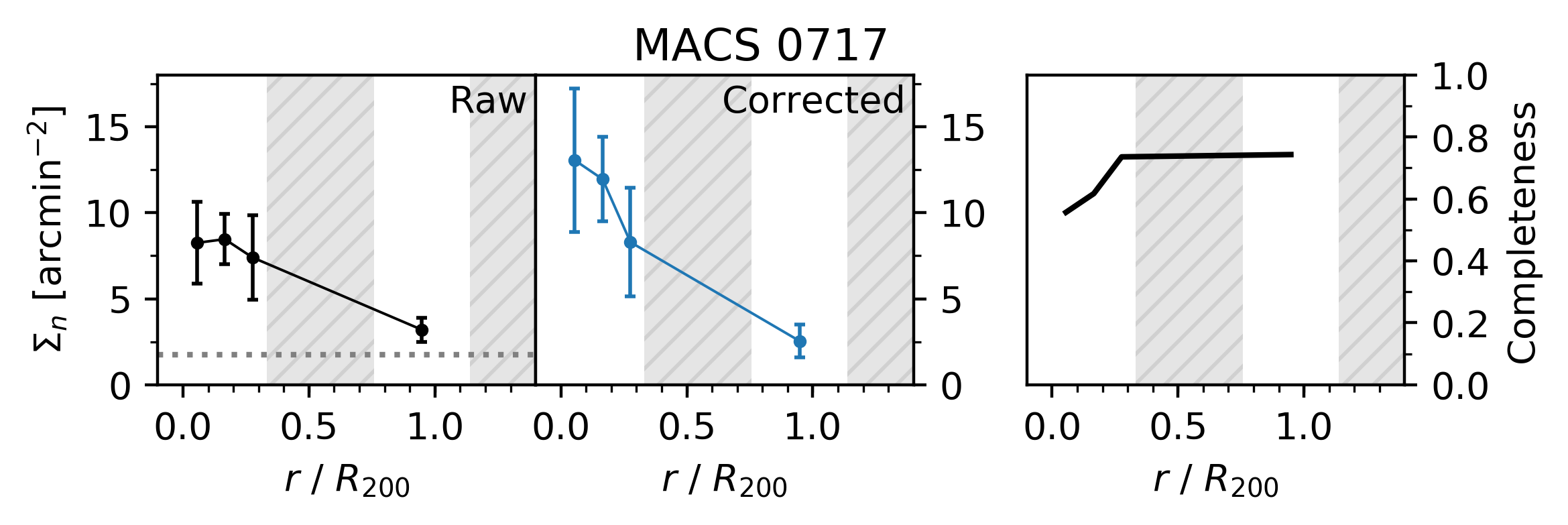}
	\caption{
    Radial surface number density profiles of UDGs in the six FF clusters.
    For each cluster, the points in the leftmost panels are the raw observed surface
    densities and the dotted line is the density of background sources
    estimated from the XDF.
    The rightmost panels shows the completeness fractions in each bin
    determined from our image simulations (see text for details).
    The middle panels show the radial profiles after correcting for
    completeness and subtracting off the estimated background contamination.
    The hatched shaded regions denote radii with no coverage (e.g.\ between the
    cluster and parallel fields).
    \label{fig:radial-profiles}
	}
\end{figure*}

Radial profiles of the surface density of UDGs were made for each of the six
clusters and are shown in Figure \ref{fig:radial-profiles}. The location of
the BCG was adopted as the cluster centre, or in the case of multiple BCGs,
their midpoint; these are listed in Table \ref{tab:clusters}.  For each
cluster, the leftmost panels in Figure \ref{fig:radial-profiles} show the raw
observed radial densities (black points) along with the density of background
sources estimated from the number of XDF sources that satisfy each cluster's
UDG cut (gray dotted line, roughly ${\sim}2~\mathrm{arcmin}^{-2}$).
The central panels shows the profile after correcting
for the completeness in each radial bin (right panels) and subtracting off the
background density.
In all six clusters, either a central depletion or a flattening out of the
radial profile of UDGs is observed.
This behaviour has been described in several nearby clusters
\citep[e.g.][]{vdb2016, mancerapina2018} and is thought to be caused by the
tidal disruption of UDGs near the centres of galaxy clusters.
However, using simulations, \cite{sales2019} find that the surface density of
UDGs rises continually towards the centre of a Virgo-like cluster.
At radii inside ${\sim}0.4 \times R_{200}$, ``tidal UDGs", a population of
galaxies transformed into UDGs as a result of tidal stripping, begin to
dominate the population, as UDGs that fell into the cluster as UDGs are now
destroyed.
Finally, it should also be noted that the highly disturbed nature of the
FF clusters renders the choice of cluster centre rather
uncertain.

\subsection{The abundance of ultra-diffuse galaxies}

\begin{figure}
	\includegraphics[width=0.50\textwidth]{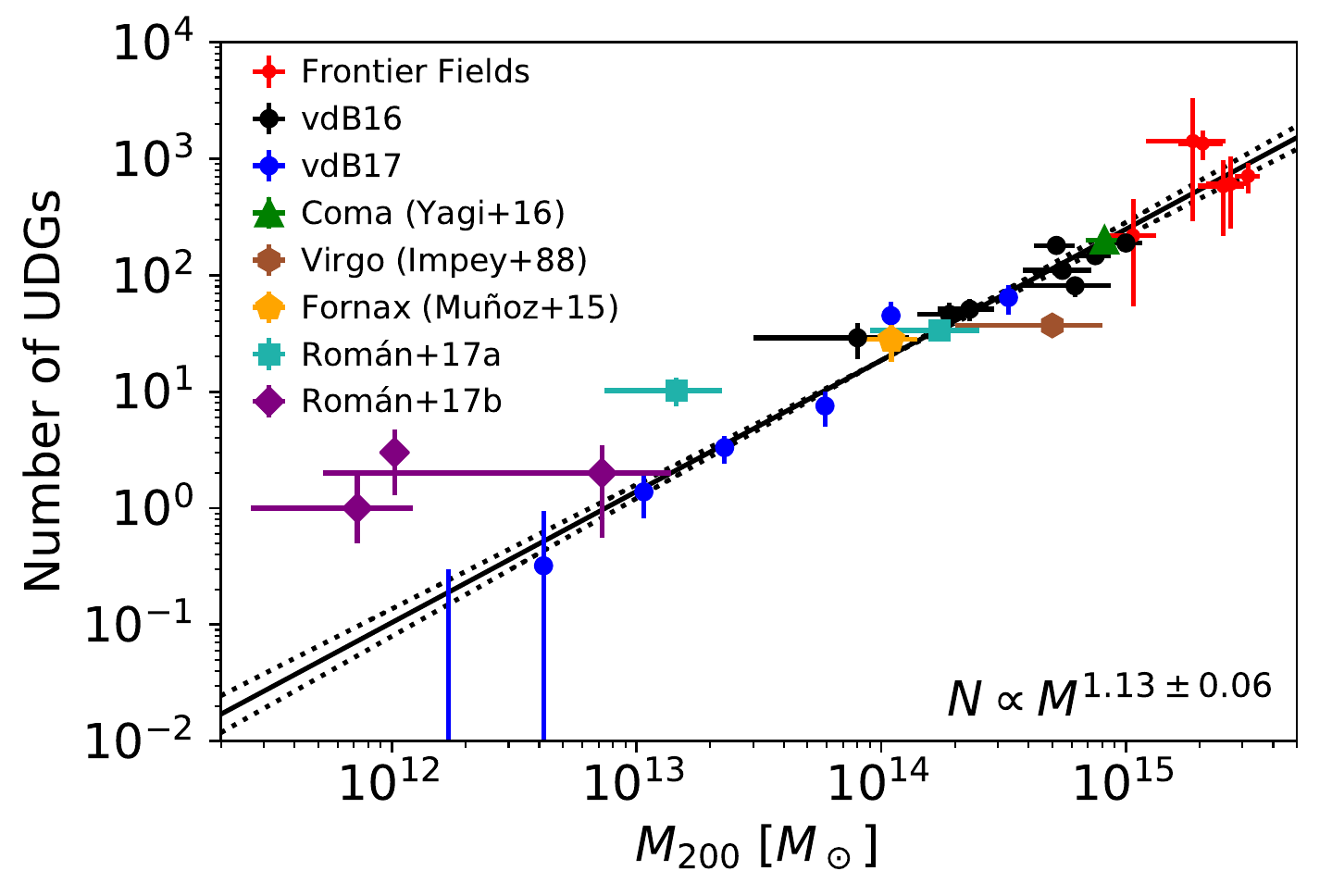}
	\caption{
    Abundance of UDGs as a function of halo mass for the six FF clusters
    examined in this work, as well as other systems from the literature (see
    text for details).
    The solid line is the best fit power law to the abundances in the FFs as well
    as the binned groups from \cite{vdb2017} and the clusters from
    \cite{vdb2016}.
    \label{fig:abundance}
	}
\end{figure}

An estimate for the total number of UDGs in each cluster is made by
integrating the corrected surface number density profile out to $R_{200}$.
These total abundances are listed in the last column of Table
\ref{tab:clusters} and range from ${\sim}200$ in MACS~0416 to ${\sim}1400$ in
Abell~2744 and Abell~S1063.
Note that for Abell~2744, Abell~S1063 and Abell~370, where we have to extrapolate
out to $R_{200}$, we assume the surface density observed in the parallel field
is constant out to $R_{200}$.
The upper and lower estimates were obtained by integrating along the lower and
upper error bars of the corrected profile, out to the upper and lower bounds
of $R_{200}$, respectively.
The large range in the abundance of UDGs in Abell~S1063 is a result of
its $R_{200}$ value being poorly constrained.

Our new estimate for the abundance of UDGs in Abell~2744 of
$1351^{+387}_{-379}$ is slightly lower than, but consistent with, our previous
result of $1961 \pm 577$ \citep{janssens2017}. This is due to our revised UDG
selection, most notably that UDGs in Abell~2744 fainter than the 50\%
completeness limit of
$\langle\mu\rangle_{e,F814W} = 26.9~\mathrm{mag}~\mathrm{arcsec}^{-2}$
were excluded from the estimate in this analysis.
The mass of Abell~2744 has been slightly lowered in this analysis as well. In
\cite{janssens2017}, an ensemble mass estimate from lensing and dynamical
studies was used, but here we only use a lensing mass estimate to be
consistent with the other five clusters studied.

In Figure \ref{fig:abundance}, we update the UDG abundance halo-mass relation
including all six FF clusters, along with the abundances in other systems from
the literature.
\cite{vdb2016} investigated the UDG populations in eight clusters at
redshifts $0.044 < z < 0.063$.
\cite{vdb2017} extended this investigation to lower masses, looking at 325
galaxy groups from the GAMA survey in seven mass bins.
For Coma, we apply our UDG criteria to the \cite{yagi2016}
catalog finding ${\sim}200$ such objects.
And similarly for Fornax, we apply our selection criteria to the
\cite{munoz2015} catalog. This catalog, however, only covers the inner 350 kpc
so we apply a geometrical correction by assuming the flat radial surface
density profile they found for dwarfs applies to UDGs out to $R_{200} =
700~\mathrm{kpc}$ \citep{drinkwater2001} and we estimate a total of ${\sim}30$
UDGs.
Finally, we include the abundances of UDGs in Abell~168 and UDG~842
\citep{roman2017a} and three Hickson Compact Groups \citep{roman2017b}.

In the Virgo cluster, \cite{impey1988} identify 27 large low surface
brightness galaxies.
Most have scale lengths $h \gtrsim 10\arcsec$ and central surface
brightnesses in the $B$-band fainter than $\mu_{0,B} \approx
23~\mathrm{mag}~\mathrm{arcsec}^{-2}$.
For $n=1$ S\'{e}rsic profiles, $\langle\mu\rangle_e = \mu_0 + 1.12$ and the
effective radius is related to the scale length via $R_e = 1.678h$
\citep{graham2005}, meaning they have $R_e \gtrsim 1.5$ kpc\footnote{Adopting
a distance of 16 Mpc to the Virgo cluster.} and
$\langle\mu\rangle_{e,B} \gtrsim 24~\mathrm{mag}~\mathrm{arcsec}^{-2}$,
satisfying the rough definition of a UDG.
In their Figure 7b, we find a total of 37 galaxies that satisfy these cuts,
including an additional 19 objects they include from \cite{caldwell1983}.

\cite{vdb2017} fit a power law to the UDG abundance halo-mass relation and
found $N_\mathrm{UDG} \propto M_{200}^{1.11 \pm 0.07}$ using the abundances in
325 galaxy groups at $0.01 \leq z \leq 0.10$ in seven mass bins and the abundances
in eight clusters at redshifts $0.044 < z < 0.063$ from \cite{vdb2016}.
The six FF clusters are ${\sim}0.5$ dex higher in mass than the most massive
system investigated in \cite{vdb2016}.
To see what effect these massive systems have on the abundance halo-mass
relation, we reperformed the fit including these new clusters. 
Orthogonal distance regression was used to fit the power law, allowing the fit
to account for uncertainty in both $N_\mathrm{UDG}$ and $M_{200}$.
We find a best fit relation of
\begin{equation}
N_\mathrm{UDG} = (19 \pm 2) \left[\frac{M_{200}}{10^{14}~M_\odot}\right]^{1.13 \pm 0.06},
\end{equation}
showing that the abundance of UDGs in these extremely massive systems at
intermediate redshift is in excellent agreement with the relation from
\cite{vdb2017} describing more local and less massive clusters.

The slope of this relation is interesting for its implications regarding the
environments where UDGs may be preferentially created or destroyed.
A slope of unity means that as structures hierarchically merge, the number of
UDGs is conserved.
A slope less than one would suggest that UDGs are more easily created in group
environments, or that they may be preferentially destroyed in clusters
\citep{roman2017b}.
On the other hand, a slope greater than unity, as we find, suggests that
the group environment is destructive to UDGs, or that there is a process that
can preferentially create UDGs in cluster environments \citep{roman2017b,
vdb2017}, with tides having already been discussed as one possibility \citep{sales2019}.

The morphology of galaxies residing in clusters changes
dramatically between $z \sim 0.4$ and $z = 0$ \citetext{i.e.\ the
Butcher-Oemler effect, \citealp{butcher1978, butcher1984, dressler1994,
moore1998}}.
The amount of light in the ICL has also grown by a factor of 2--4 since
$z \sim 1$ \citep{burke2012}, and the tidal disruption of galaxies is
understood to be its origin \citep[e.g.][]{harris2017}.
This suggests there is a big caveat to including the abundances of UDGs in both local and
intermediate redshift systems on the same abundance halo-mass relation as it
is not unreasonable to believe that this relation has
evolved since $z \sim 0.4$, with these processes destroying (or creating) UDGs
as clusters have continued to assemble over the past ${\sim}4$ Gyr.
That said, the abundance of UDGs in these extremely massive intermediate redshift
clusters agrees very well with the relation at $z \sim 0$.

\subsection{Spatial distributions}\label{sec:spatialdist}

We now return to the remarkable projected spatial distributions of UDGs that
we first drew attention to in \S \ref{sec:udgselection}.
Maps were made to investigate the spatial distributions of UDGs and UCDs and
their relation to other structures in the clusters.
These are shown in Figure \ref{fig:maps}.
We restrict ourselves to the WFC3 region (pink outline) as this is where our
UDG selection was performed.
The red points are the locations of UDGs while UCDs are marked in black.
The blue triangles mark the positions of other cluster galaxies, selected with
$R_{e,c} \geq 0.5~\mathrm{kpc}$ and $|z_\mathrm{phot} - z_\mathrm{cl}| \leq
0.05$, where $z_\mathrm{cl}$ is the cluster redshift.
Black text labels mark the positions of mass peaks from gravitational lensing
analyses in the literature, which we include to see whether there is any possible
relationship between substructures and concentrations of UDGs.
The red-yellow contours are smoothed \textit{Chandra} X-ray fluxes
(ObsIDs 8477, 18611, 515, 16304, 16306 and 16305) tracing the hot
intracluster medium (ICM).
The blue-green contours are mass surface density contours from the Merten
gravitational lensing models\footnote{Obtained from
\url{http://www.stsci.edu/hst/campaigns/frontier-fields/Lensing-Models}.}
\citep{merten2009,merten2011,zitrin2009,zitrin2013}. 
The X's mark different possible definitions of the cluster centres: in
blue is the peak mass surface density, in pink is the BCG centre (Table
\ref{tab:clusters}) and in red is the centroid of the UDG distribution.
Multi-lobed mass distributions and/or any disagreement between the mass
contours, ICM contours, and the BCGs point towards clusters still in the
process of being assembled. 

\begin{figure*}
	\includegraphics[width=0.50\textwidth]{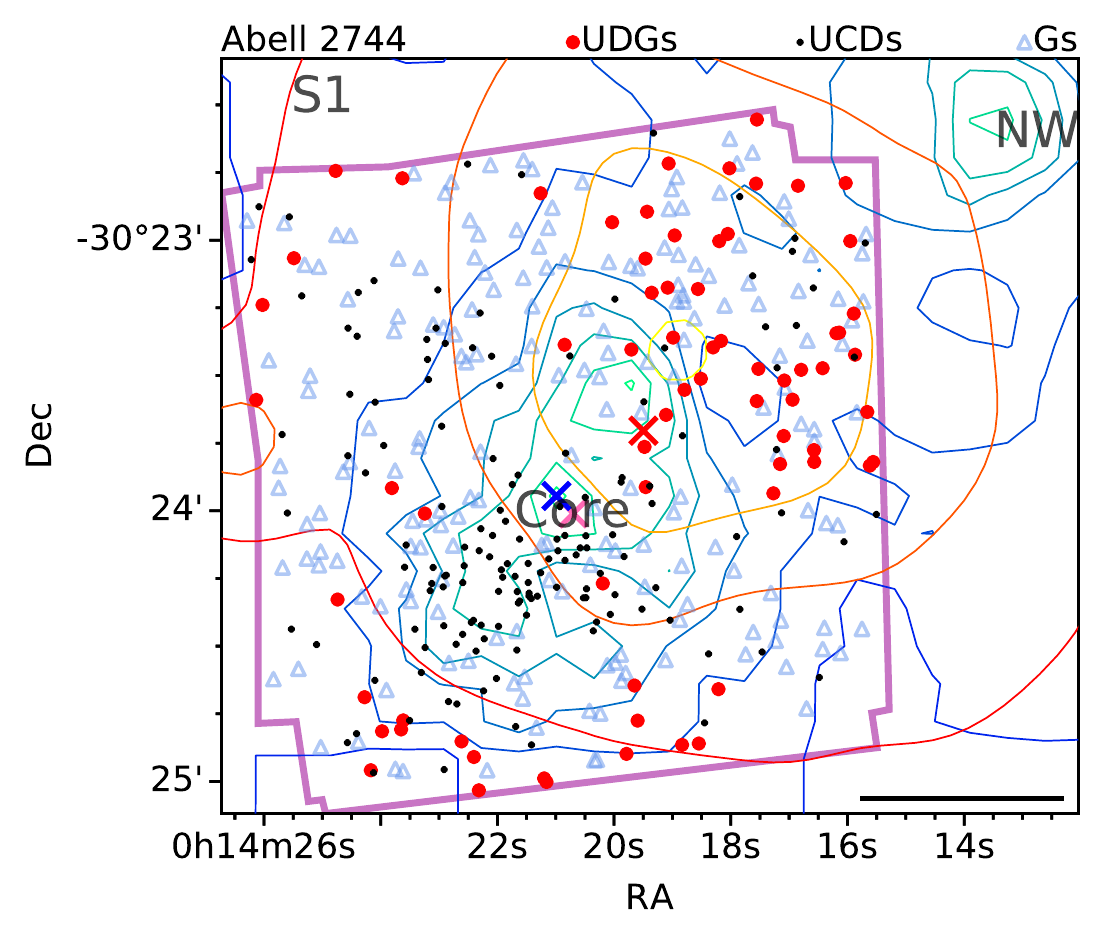}
    \includegraphics[width=0.50\textwidth]{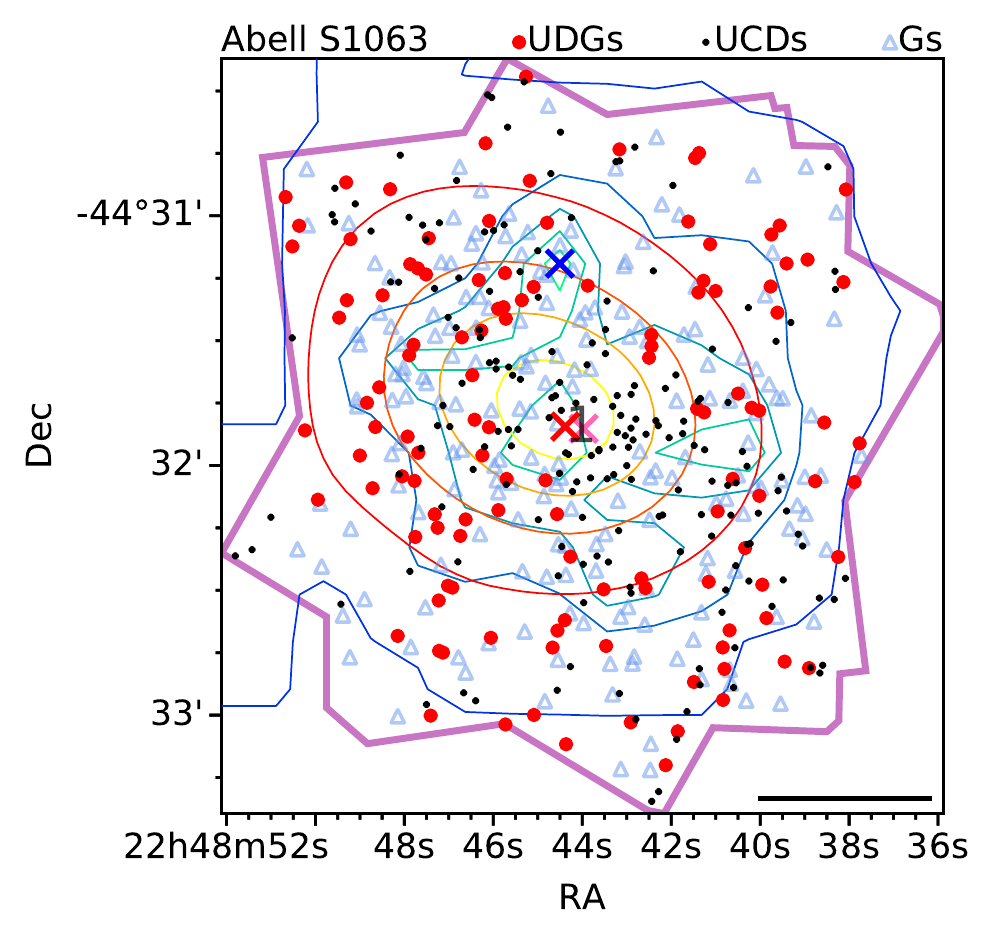}
	\caption{
    Maps showing the projected spatial distributions of various components of the
    clusters Abell~2744 and Abell~S1063 within the WFC3 cluster fields (pink
    border).
    The black line in the lower right measures 200 kpc in length.
    Locations of UCDs are marked in black, UDGs in red, and other cluster
    galaxies (labelled Gs) are marked in blue.
    The X's mark different possible adoptions of the cluster centre: pink, the
    BCG centre (that listed in Table \ref{tab:clusters}); blue, the
    gravitational lensing mass surface density peak; and red, the mean UDG
    location.
    The blue-green contours trace the gravitational lensing surface mass
    density of the cluster
    \citep{merten2009,merten2011,zitrin2009,zitrin2013}, while the red-yellow
    contours are Chandra X-ray fluxes.
    Labels in black correspond to cluster substructures from the literature.
    In Abell~2744, `Core', `NW' and `S1' are the three substructures
    identified by \cite{jauzac2016} that are in proximity to the WFC3 field. 
    The single cluster-scale mass component coincident with the BCG found by
    \cite{richard2014} is marked `1' in Abell~S1063.
    UDGs appear deficient in the most dense cluster environments, with UCDs
    instead being abundant towards the cluster centres.
    No relationship between UDG locations and the X-ray flux (tracing the hot
    ICM) is observed, as is expected if UDGs are gas poor
    systems.
    \label{fig:maps}
	}
\end{figure*}
\begin{figure*}
    \ContinuedFloat
	\includegraphics[width=0.50\textwidth]{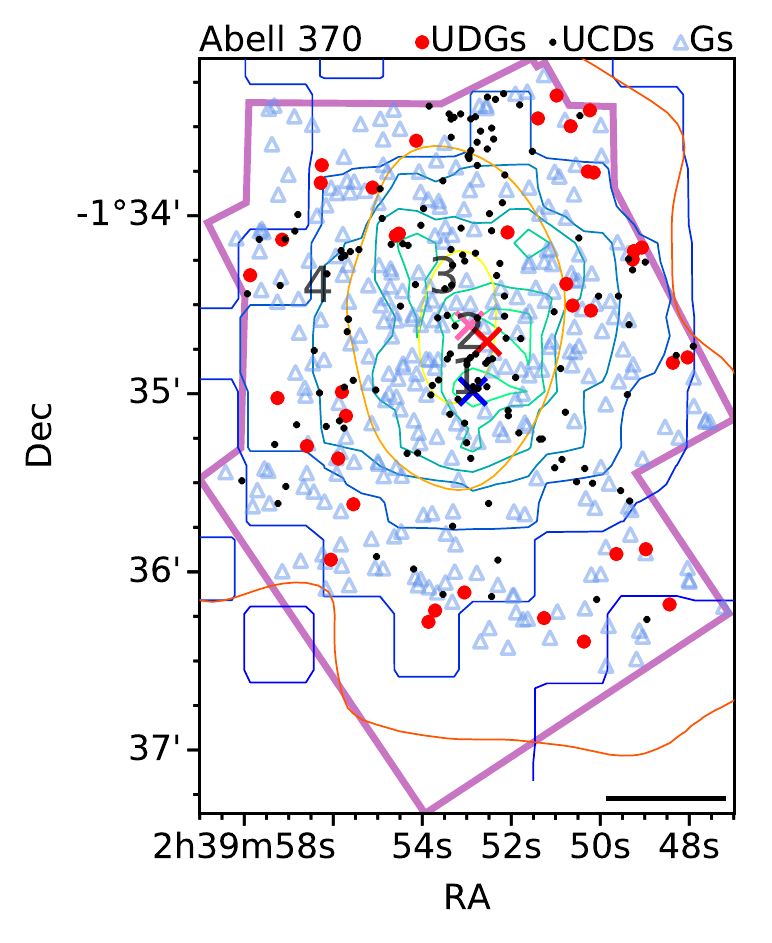}
	\includegraphics[width=0.50\textwidth]{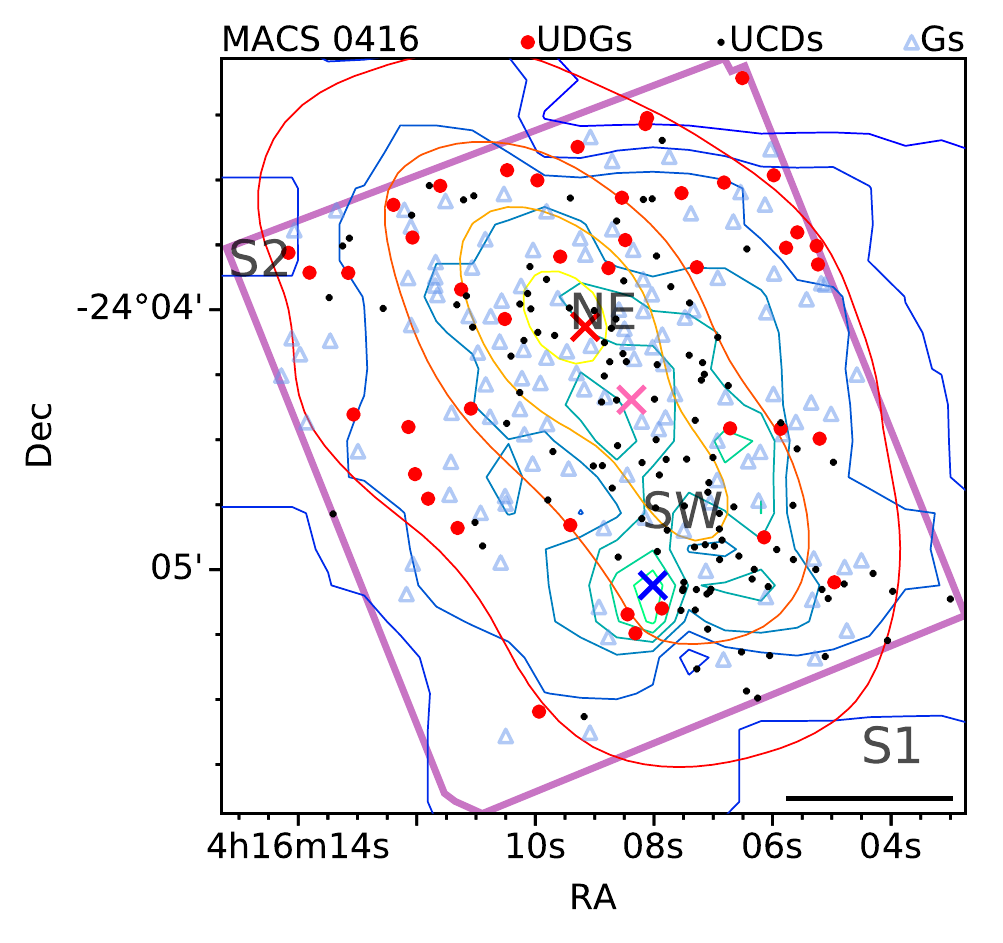}
    \caption{
    (Continued.) Abell~370 and MACS~0416.
    In Abell~370, 1--4 mark the positions of DM1--DM4, the four `large-scale'
    mass components identified by \cite{lagattuta2019}.
    In MACS~0416, NE and SW mark the positions of the two main dark matter
    halos comprising the core of MACS~0416, while S1 and S2 are two additional
    galaxy group sized (${\sim}10^{13}~M_{\odot}$) substructures, all from
    \cite{jauzac2015}.
    }
\end{figure*}
\begin{figure*}
    \ContinuedFloat
	\includegraphics[width=0.50\textwidth]{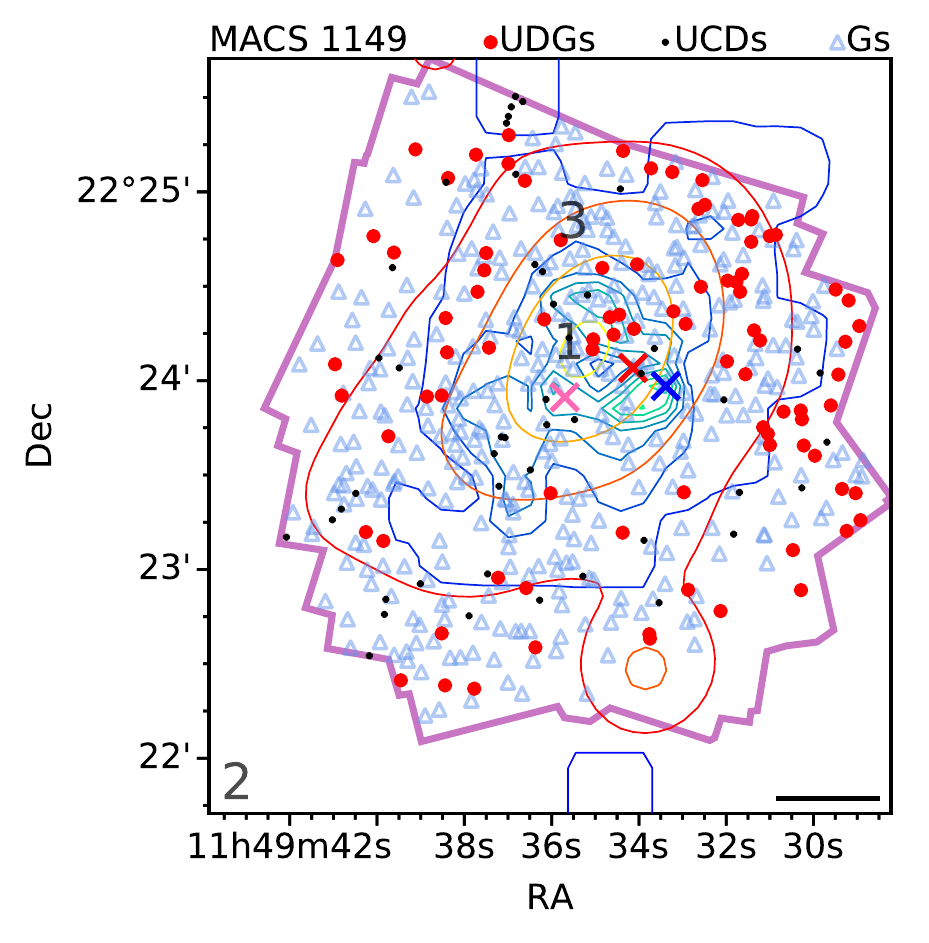}
	\includegraphics[width=0.50\textwidth]{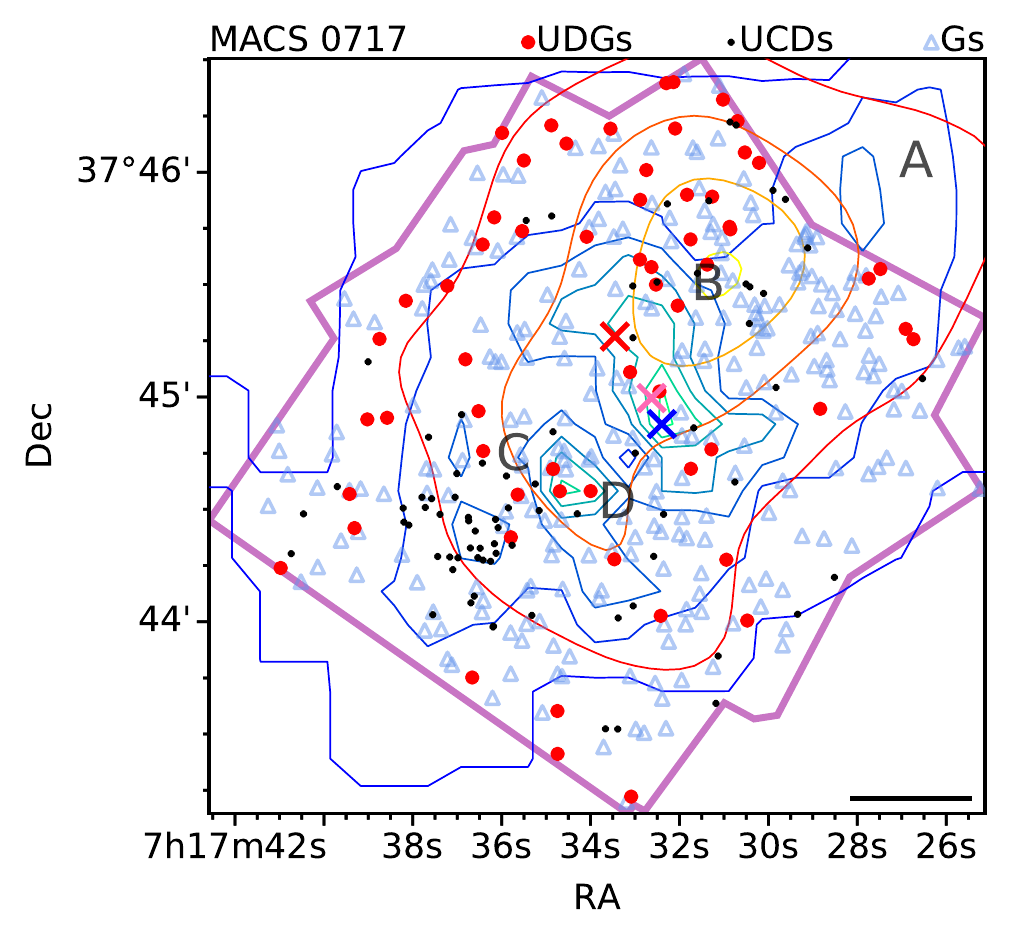}
    \caption{
    (Continued.) MACS~1149 and MACS~0717.
    In MACS~1149, the positions of the three subclusters identified by
    \cite{golovich2016} are marked 1, 2 and 3.
    In MACS~0717, A, B, C and D mark the \cite{limousin2016} NFW fit positions
    of the \cite{ma2009} light peaks in the cluster core.
    }
\end{figure*}

\begin{figure*}
	\includegraphics[width=0.50\textwidth]{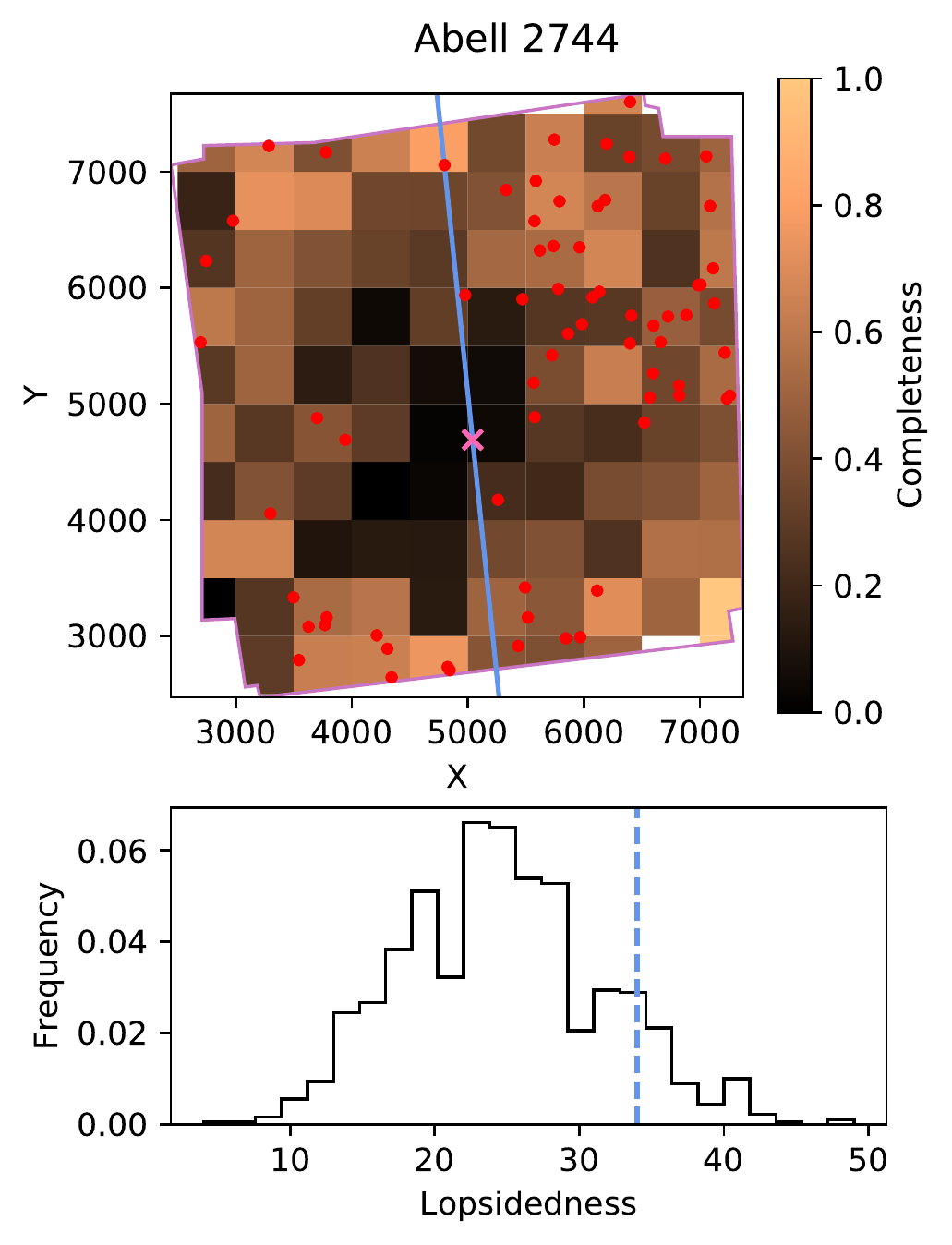}
	\includegraphics[width=0.50\textwidth]{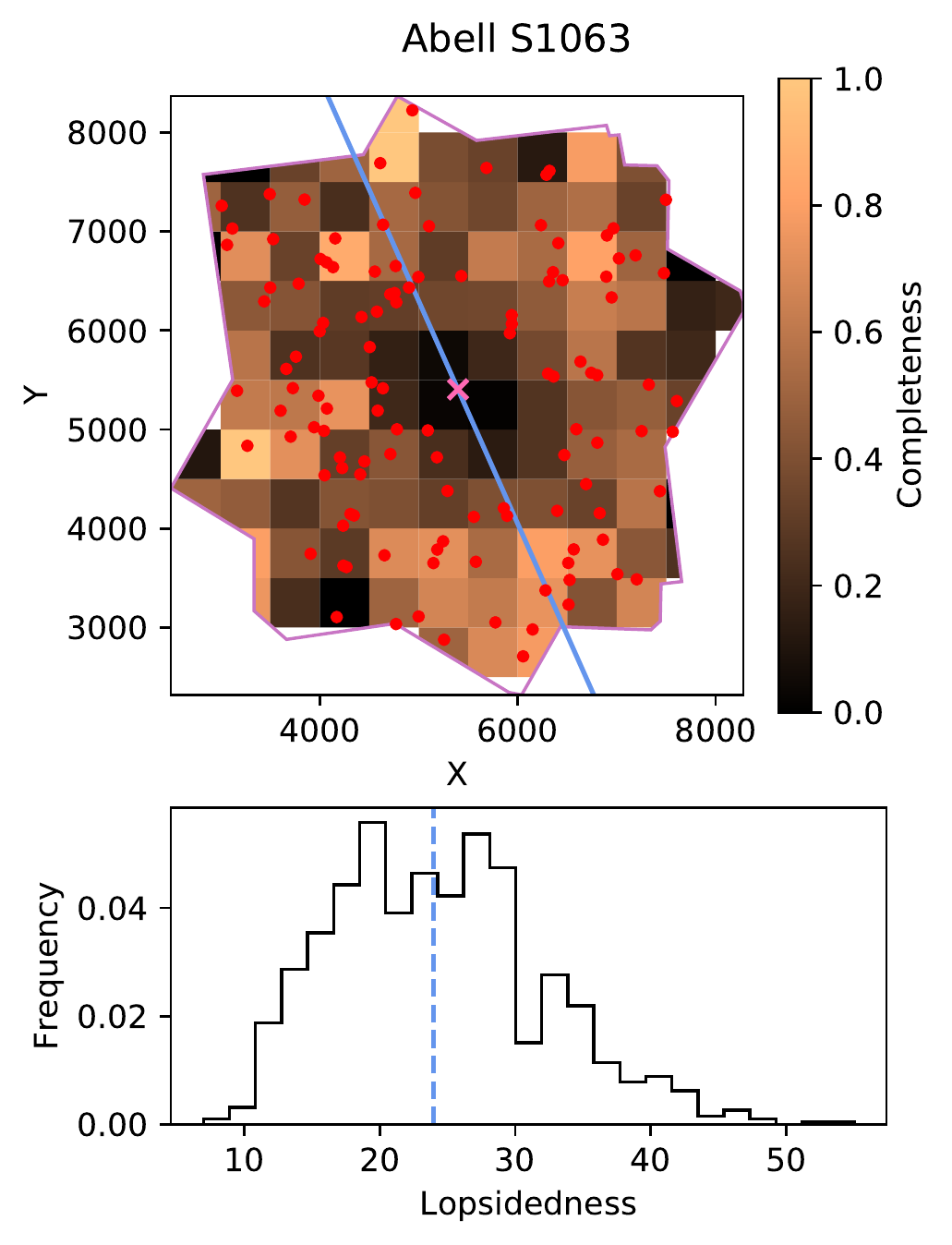}
	\caption{
    A simple test for a lopsided projected spatial distribution of UDGs in the
    cluster core fields. Top: We search for the straight line that passes through
    the adopted cluster centre (pink cross, see Table \ref{tab:clusters})
    which results in the most UDGs (red points) being on one side. The
    ACS/WFC3 overlap region is divided into 500 pixel $\times$ 500 pixel cells
    and the image simulations are used to determine a completeness fraction in
    each.  The pink border is the WFC3 $F160W$ footprint.  Blank regions
    inside the WFC3 footprint are either small slices of cells that had no
    injected sources in the image simulations or are regions where the WFC3
    footprint extends beyond the ACS coverage.
    Bottom: To test whether this is a real phenomenon or a completenes
    artifact, for each cluster, we create 1000 realizations of a uniform
    spatial distribution of UDGs. An estimate of the completeness corrected
    total number of UDGs, $N_\mathrm{cor}$, in the core field is made by dividing the
    number of UDGs in each cell by its completeness fraction.  $N_\mathrm{cor}$
    locations are then randomly drawn, with each location having a probability
    of being kept equal to the completeness in that cell. We again search for
    the line passing through the centre that results in the most lopsided
    configuration. The distribution of resulting lopsidedness (defined as the
    difference between the more and less populated sides) is shown, with the
    observed lopsidedness denoted with the dashed line.
    This phenomenon is least pronounced in Abell~S1063, and this is the most
    relaxed FF cluster \citep{lotz2017}. 
    \label{fig:lopsidedness}
	}
\end{figure*}
\begin{figure*}
    \ContinuedFloat
	\includegraphics[width=0.50\textwidth]{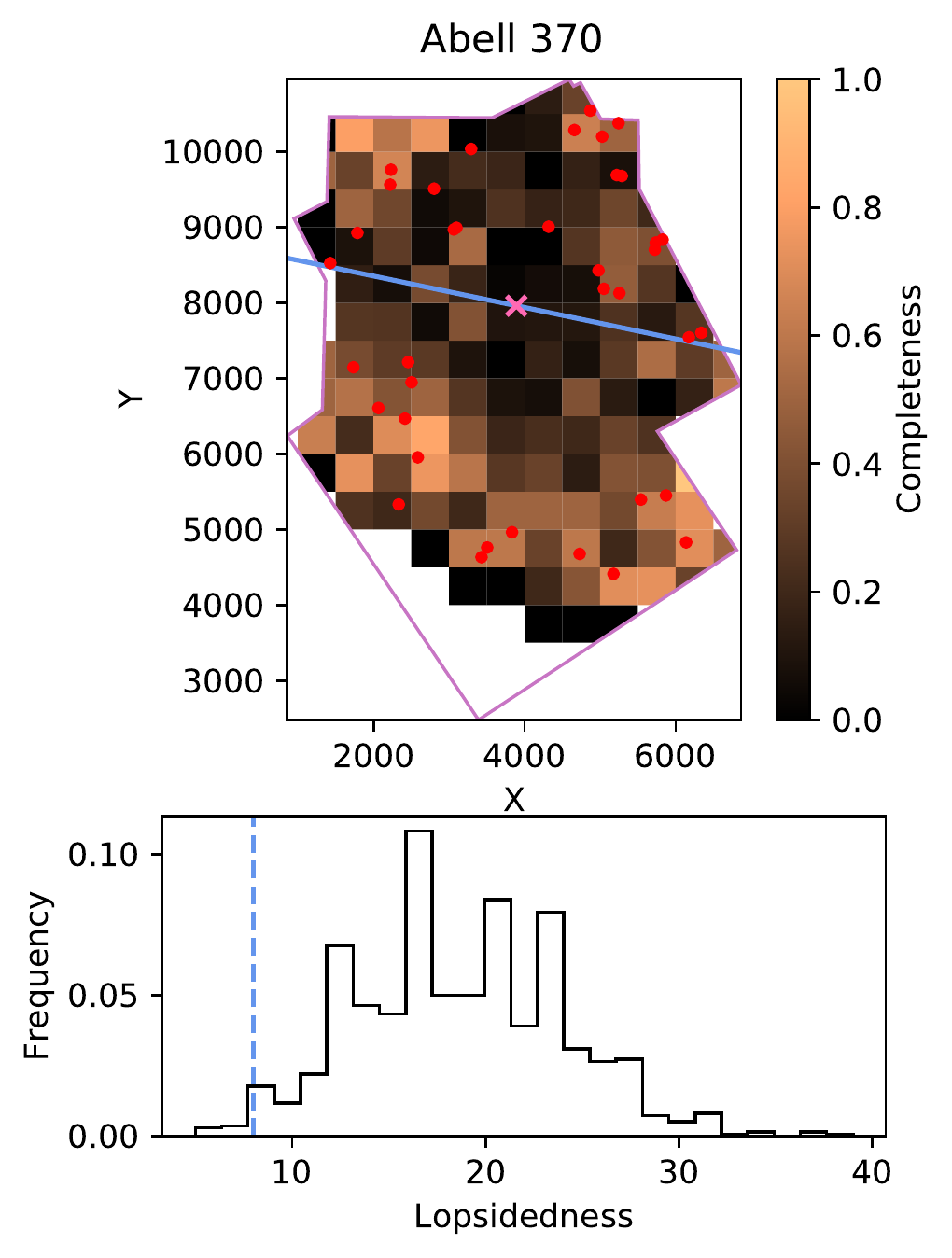}
	\includegraphics[width=0.50\textwidth]{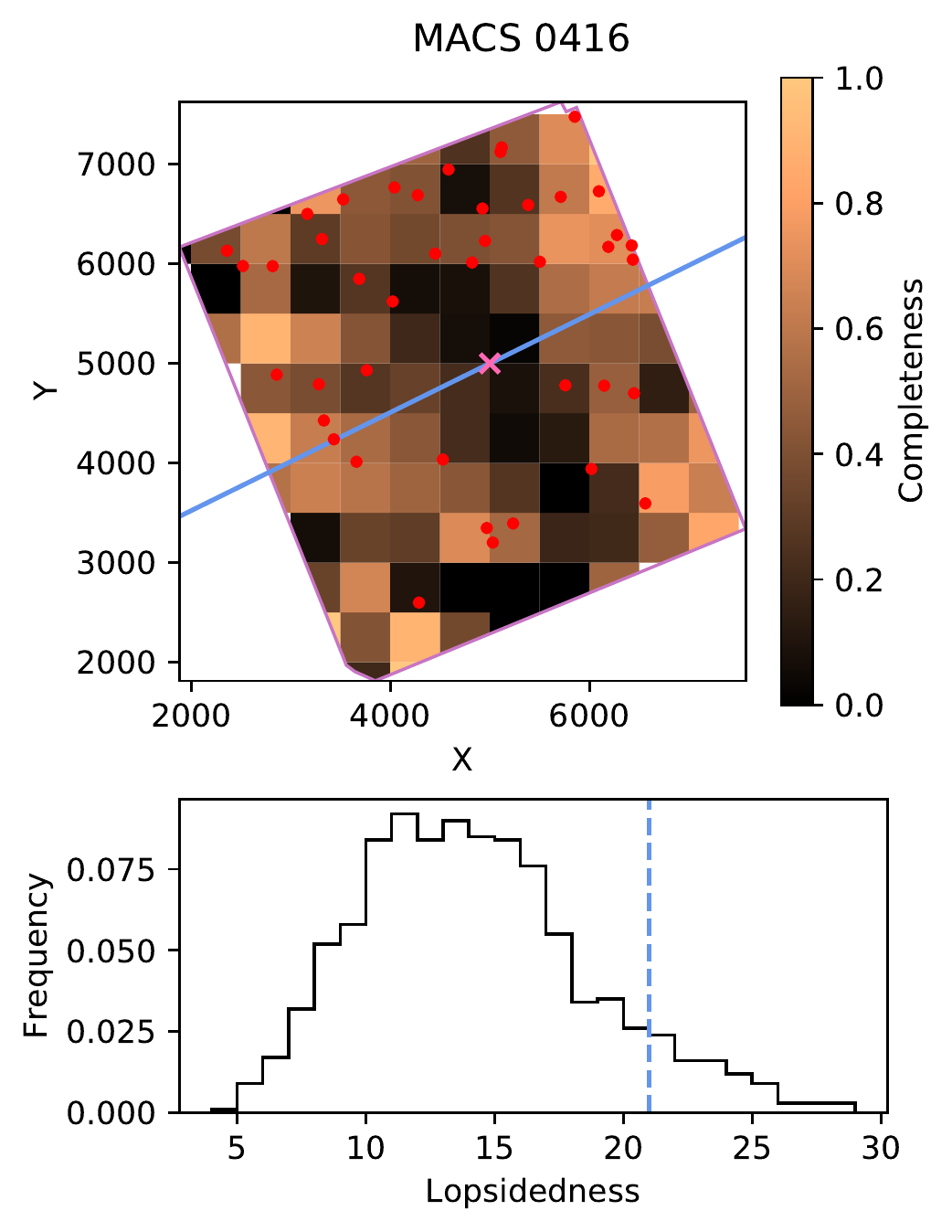}
    \caption{(Continued.)}
\end{figure*}
\begin{figure*}
    \ContinuedFloat
	\includegraphics[width=0.50\textwidth]{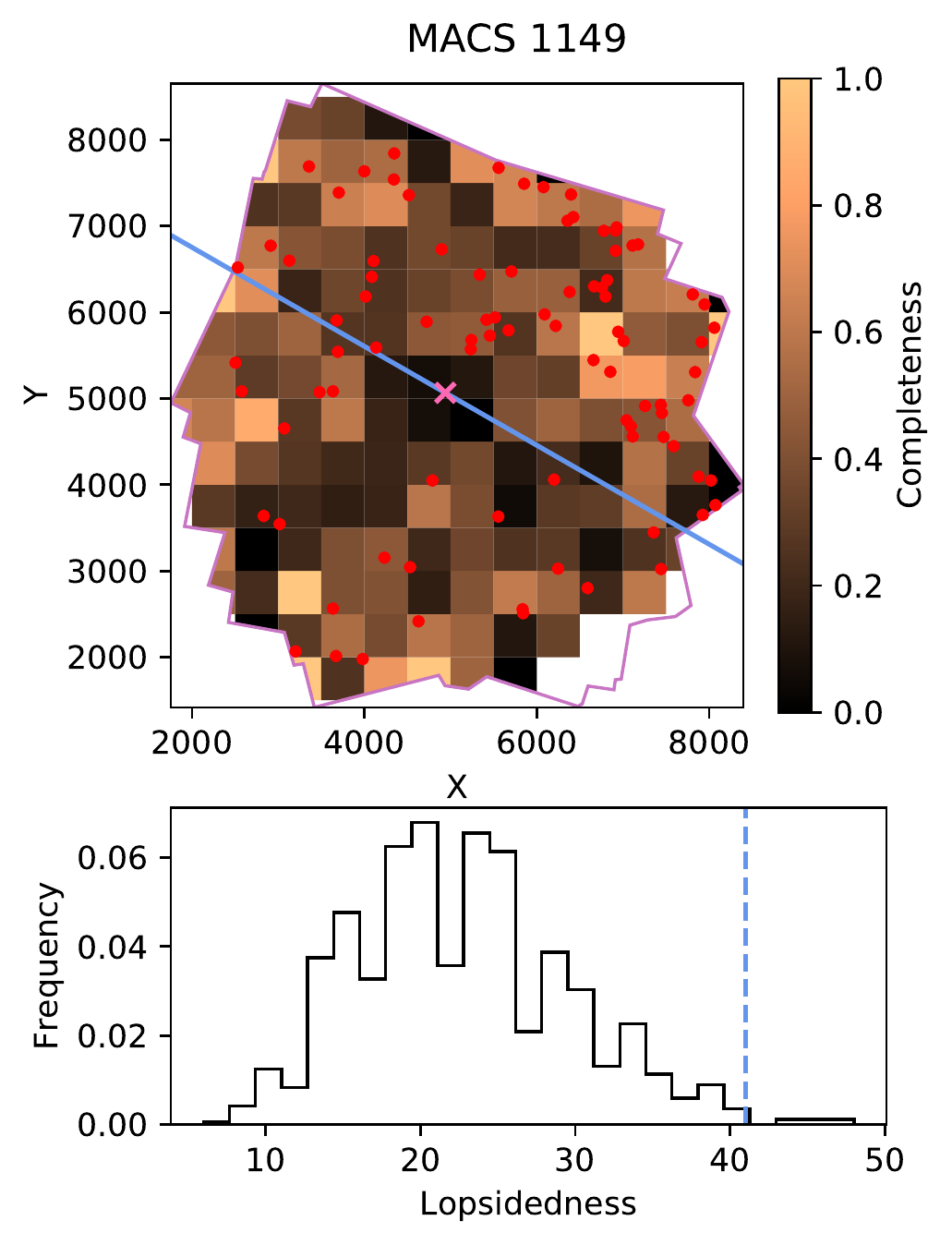}
	\includegraphics[width=0.50\textwidth]{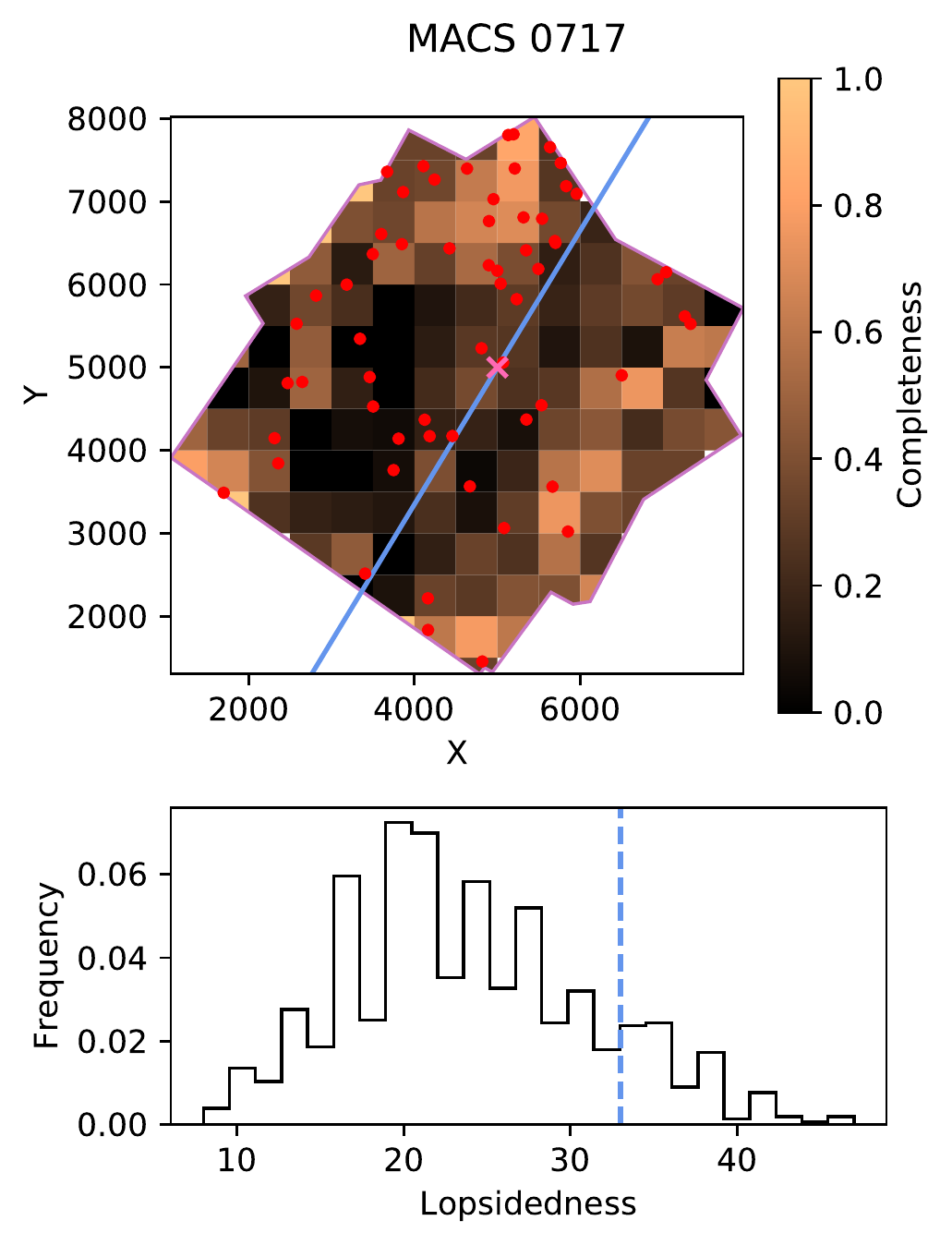}
    \caption{(Continued.)}
\end{figure*}

In four of the six clusters, the spatial distribution of UDGs does not appear
uniform, with many more UDGs on one side than the other.
Only Abell~S1063 and Abell~370 present with what appears to be an azimuthally
even distribution of UDGs around the cluster.
Given the highly disturbed nature of these clusters, this may be expected.
However, this behaviour is not seen amongst the other cluster galaxies.
We are now in a position to try and understand if this effect is real.

A simple measure of this lopsidedness was devised by finding the line that bisects
the cluster which results in the most UDGs being on one side.
This line is shown in the top panels of Figure \ref{fig:lopsidedness}, with the
positions of UDGs also marked.
We also wanted to test whether the uneven distributions were simply a
reflection of the completeness fraction across the clusters.
To this end, the WFC3 regions of each cluster were split into cells and the
completeness fraction was computed in each using our image simulations; this
is shown as the background of the top panels of Figure \ref{fig:lopsidedness}.
A rough estimate of the completeness corrected number of UDGs,
$N_\mathrm{cor}$, was made by
dividing the number of UDGs in each cell by the cell's completeness fraction
and summing.
We then performed 1000 simulations of a uniform spatial distribution of UDGs
by randomly drawing $N_\mathrm{cor}$ positions and then simulating the
completeness by giving each position a probability of being kept equal to the
completeness in its cell.
The results of these simulations are shown in the bottom panels of Figure
\ref{fig:lopsidedness}.
Defining lopsidedness as the difference between the more and less populated
sides, the histogram shows the distribution of lopsidedness from the
simulations and the dashed blue vertical line is the observed value.
All but Abell~S1063 show spatial distributions of UDGs inconsistent with that
expected from a uniform distribution based on our simulations.
In the case of Abell~370, the detected UDGs are quite evenly distributed on
either side, but based on the completeness, this should not be the case. If the
true UDG distribution were uniform, we would expect to detect $\sim$15--25
more UDGs in the southern half of the image.

We then performed the same test on the UDGs in the parallel fields.
These are shown in Appendix \ref{sec:lopsidedness-parallel} and Figure
\ref{fig:lopsidedness-parallel}.
In the parallel fields, the distribution appears much more uniform and only one
field, the Abell~370 parallel field, shows any evidence of a possible lopsided UDG
distribution in the simulations.

We now briefly discuss each of the clusters in turn.

\paragraph{Abell~2744} Abell~2744 is a massive complex merging cluster with
past and ongoing mergers between at least four substructures \citep{owers2011,
merten2011, medezinski2016} and as many as eight \citep{jauzac2016}.
The bulk of the UDGs are located in the
northwest quadrant of the field, northwest of the cluster core.
This cloud of UDGs is located roughly mid-way between the cluster core and a
relatively massive NW substructure consistently found in all analyses located ${\sim}580$ kpc
northwest of the cluster centre \citep{merten2011, medezinski2016,
jauzac2016}.
Three of the eight substructures identified by \cite{jauzac2016} that are in
proximity to the WFC3 field, including the core and NW substructures, are
marked in Figure \ref{fig:maps}.
X-ray observations of the NW substructure reveal a trail of cool gas to the south
and a cold front to the north, suggesting this substructure is moving
northward on its first infall \citep{jauzac2016}.
The other merger of possible relevance to our UDGs has already occurred and
was the passage of the Northern substructure through the cluster core
\citep{owers2011, merten2011, medezinski2016}.
It is possible this collection of UDGs was deposited here in the process of
the past north-south merger, or alternatively they are possibly associated
with the NW subcluster on its first infall.
The UCDs in this cluster are heavily concentrated around the three BCGs
southeast of the core mass peak.

\paragraph{Abell~S1063} Abell~S1063 possesses one of the highest known X-ray
temperatures and is possibly undergoing a major merger, with the merger axis in the plane
of the sky \citep{gomez2012}.
But despite this possible merger, Abell~S1063 is the most relaxed FF cluster
\citep{lotz2017}, with the smoothest mass contours \citep{gruen2013,
diego2016}.
In contrast to the other FF clusters, \cite{richard2014} only require a single
cluster-scale dark matter component to fit the observed lensed images in
Abell~S1063; the centre of this `DM1' component is coincident with the BCG and
is marked `1' in Figure \ref{fig:maps}.
However, in addition to this central halo, \cite{johnson2014} find two other
cluster-scale halos, one ${\sim}400\arcsec$ (2 Mpc) to the northeast, and the
other ${\sim}100\arcsec$ (500 kpc) to the south, but both are well outside the
\textit{HST} field of view.
Abell~S1063 presents the most uniform distribution of UDGs azimuthally around
its centre, with the central region deficient in UDGs but abundant in UCDs.

\paragraph{Abell~370} Abell~370 is a massive merger of two roughly equal
subclusters along the line of sight, with each BCG belonging to one of the
subclusters \citep{richard2010}.
However, while the northern BCG has a slightly higher redshift than its southern counterpart,
\cite{lagattuta2019} find only a single peak in the redshift distribution of
the cluster members, suggesting that the merger is either in the plane of the
sky or has already taken place.
In their best-fit `\textit{copper}-class' model, \cite{lagattuta2019} identify
four large-scale massive components (DM1--DM4), whose positions are marked
1--4 in Figure \ref{fig:maps}, in addition to a handful of smaller galaxy-scale components. 
DM1 and DM3 correspond to the mass clumps associated with the southern and
northern BCGs, respectively. DM2 is a `bar' between the two BCGs, and DM4 is
associated with a `crown' of galaxies in the northern portion of the field.
As discussed above, based on the completeness, the northern region of the
Abell~370 WFC3 field is overabundant in UDGs.
The UCDs are fairly evenly spread amongst the cluster ellipticals.

There is also the presence of the bright foreground elliptical galaxy
PGC~175370 (distance ${\sim}200$ Mpc) on the northern edge of the ACS image to
consider (see Figure \ref{fig:locations}).
It cannot be ruled out that some of the `extra' UDGs found in the northern
portion of Abell~370 based on our completeness simulations may instead be
`regular' dwarf galaxies associated with PGC~175370.
The group of UCD candidates at the northern edge of the WFC3 coverage are
coincident with a cluster elliptical (compare with Figures \ref{fig:locations}
and \ref{fig:pgc175370}).
On this basis we believe them to be genuine UCD candidates and not GCs
associated with PGC~175370.

\paragraph{MACS~0416} MACS~0416 is composed of two main subclusters undergoing
a merger.  Originally thought to be observed after a possible binary head-on
merger \citep{mann2012, jauzac2015}, more recent radio and X-ray observations
point toward the subclusters in MACS~0416 being observed in a pre-collisional
state \citep{ogrean2015, balestra2016}. Each of the subclusters, however, may
have been formed in a recent merger of their own \citep{balestra2016}.
In Figure \ref{fig:maps}, we mark the positions of the NE and SW subclusters from
\cite{jauzac2015}\footnote{The NE and SW subclusters are denoted C1 and C2 in
Table 2 of \cite{jauzac2015}, respectively.}, in addition to two galaxy group sized
(${\sim}10^{13}~M_{\odot}$) substructures they found near the core, S1 and S2. The motion
of the SW subcluster is towards us whereas the NE component is receding.
We detect many more UDGs around the NE subcluster than the SW, and a few in
the northeast corner may be associated with the S2 substructure.
The UCDs in MACS~0416 are concentrated along the bridge of cluster ellipticals
and ICL joining the two subclusters.

\paragraph{MACS~1149} \cite{golovich2016} identify three subclusters
comprising the core of MACS~1149 and their positions are marked in Figure
\ref{fig:maps}.
In their merger scenario, subclusters 1 and 2, with masses of ${\sim}1.7$ and
${\sim}1.1 \times 10^{15}~M_{\odot}$, respectively, have already merged and
passed through one another along a merger axis close to the plane of the sky.
Subcluster 3 is an order of magnitude less massive with a mass of ${\sim}1.2
\times 10^{14}~M_{\odot}$. Its merger with subcluster 1 is along the line of
sight and has recently taken place with the subclusters near pericentre, and
subcluster 3 now receding into the background.
Interestingly, no concentration of UCDs is seen near the BCG, and there is not
an excess of UCD candidates in the cluster field relative to the parallel
field.
MACS~1149 has the faintest BCG and the least ICL, perhaps hinting that the
processes that build these components in clusters are linked to the formation
of UCDs.

\paragraph{MACS~0717} MACS~0717 is another complex merging cluster. At
$z=0.545$, it is
the most massive cluster known at $z > 0.5$ \citep{edge2003, ebeling2004,
ebeling2007, jauzac2018}. Its
core contains four massive merging components \citep{ma2009, limousin2012},
surrounded by seven more substructures at
projected radii between 1.6 and 4.9 Mpc \citep{jauzac2018}. It also hosts a
filament extending a projected distance of ${\sim}4.5$ Mpc to the southeast,
with a true length of ${\sim}18$ Mpc, feeding mass into the cluster core from
behind \citep{ebeling2004,jauzac2012}. By $z=0.308$, it will likely be more massive
than Abell~2744, and by $z=0$, it will grow to a ${\sim}10^{16}~M_{\odot}$
supercluster \citep{jauzac2018}.

Of the four core subhalos, whose giant elliptical galaxy concentrations are
denoted A, B, C and D by \cite{ma2009}, the overdensity of UDGs we detect in
the upper portion of the WFC3 coverage appears to be possibly spatially
associated with subhalo B. 
The NFW fit positions of the mass peaks from \cite{limousin2016} associated
with these light peaks are marked in Figure \ref{fig:maps}.
Because of its undisturbed cool core, subhalo B is thought to be on its first
infall into the cluster at a relative velocity of ~3000
$\mathrm{km}~\mathrm{s}^{-1}$ \citep{ma2009}.
The UCDs, on the other hand, are concentrated near the southeast BCG.\\

In clusters with uneven spatial distributions of UDGs, we find
tentative associations between UDGs and detected substructures in
the clusters, often on their first infall.
This is in stark contrast to Coma, where the UDGs are concentrated around the cluster centre
and are thus likely longtime cluster members \citep{koda2015}.
Only in the most relaxed FF cluster do we find similar behaviour.
The fact that UDGs are found in groups and the near-linear UDG abundance
halo-mass relation discussed above suggest that UDGs are not only formed in clusters, but
are also formed outside and fall in.
Coma and Abell~S1063 demonstrate that many UDGs survive the
relaxation of substructures to become mixed throughout clusters, save the densest regions.
In their dissection of the Virgo Cluster, \cite{vcc6} find different spatial
distributions for the various morphological types.
In particular, they find that bright dwarf elliptical (dE) galaxies are more
concentrated than faint ones, and an even stronger effect is seen where
nucleated dwarfs are more concentrated than non-nucleated ones.
They also find that the radial number density profiles of dEs are well fitted
by either exponential or King profiles, with no deficiency at the centre.
Recall that many objects that perhaps would now be classified as UDGs are not a
separate morphological class in that analysis, and are instead included with
the dEs.

The ongoing Beyond Ultra-deep Frontier Fields And Legacy Observations
(BUFFALO) will triple the ACS coverage for each cluster and quadruple the WFC3
coverage.
The increased coverage will permit study of additional substructures that are just
missed by the currently available WFC3 data and nearly fill in the regions between
the cluster and parallel fields.

\subsection{UDGs as possible UCD factories?}

There is growing evidence that the centres of galaxy clusters are destructive
to UDGs, with lower mass UDGs particularly susceptible \citep[e.g.][]{vdb2016, sales2019}.
There are hints that low-mass UDGs may host particularly massive GCs
\citep[e.g.\ NGC1052-DF2,][]{vandokkum2018b} that are UCD-like in their
luminosities, but it is far too early to say if this holds for the entire
population of such systems.
Two UDGs in the Virgo cluster, VLSB-A and VLSB-D, are possibly undergoing
such a tidal disruption \citep{mihos2017, toloba2018}.
In the FFs, we find that ${\sim}15\%$ of UDGs are either nucleated or host
nearby compact systems that could survive the destruction of their hosts.
Half of all UDGs with nuclei or point sources are found in the most relaxed
cluster, Abell~S1063, with the individual fractions ranging from 4\% in
MACS~0717 to 25\% in Abell~S1063, suggesting that perhaps their formation is
tied to the relaxation of clusters.
And recall that all point sources detected in the FFs are far too luminous to
be GCs, with GCs in the nearest FF cluster having apparent magnitudes $m_\mathrm{814}
\gtrsim 31$ mag.
Nucleated dwarfs are also more concentrated in clusters \citep{vcc6},
suggesting that the formation of a nucleus is tied to denser cluster
environments.
In Coma, over 50\% of UDGs host nuclei \citep{yagi2016}.
It is possible that this fraction has evolved significantly since $z \sim
0.4$, but observational biases cannot be ruled out, with faint nuclei almost
certainly being missed at the distances to the FF clusters.

In all of the FF clusters---but especially in Abell~2744 and Abell~370---the
spatial distributions of UDGs and UCDs appear nearly opposite (see Figure
\ref{fig:maps}).
This observation was confirmed by running a 2D Kolmogorov-Smirnov test
\citep{nrc} comparing the positions of UDGs and UCDs in each cluster field,
and in each case, the differences between the spatial distributions of UDGs
and UCDs are confirmed to be statistically significant.\footnote{The highest
$p$-value is 0.03 for both Abell~370 and MACS~1149, the lowest is $4.6 \times
10^{-9}$ for Abell~2744.}

\begin{figure}
	\includegraphics[width=0.50\textwidth]{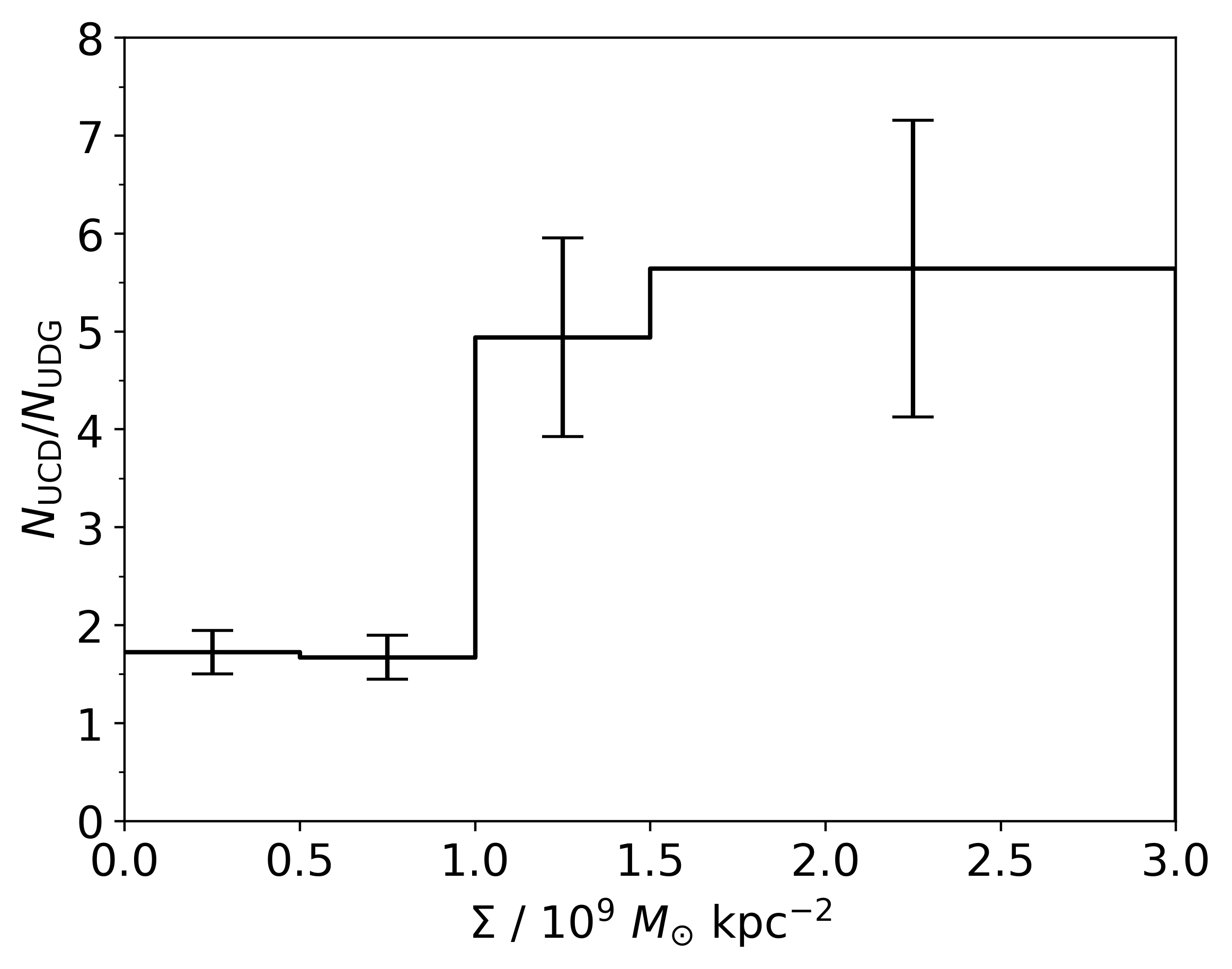}
	\caption{
    The ratio of UCDs to UDGs in bins of mass surface density across all six
    FF clusters.
    A single bin for mass surface densities greater than $1.5 \times
    10^9~M_\odot~\mathrm{kpc}^{-2}$ was used due to the small areas of the
    clusters with such high densities.
    For each UDG and UCD, we used the wide-field low-resolution Merten lensing
    models \citep{merten2009,merten2011,zitrin2009,zitrin2013} to compute the
    mass surface density at its location.
    The abundance of UCDs relative to UDGs roughly triples in the high density
    regions of the clusters.
    \label{fig:ucdudg-ratio}
	}
\end{figure}

A different attempt to quantify this phenomenon was made by looking at the
ratios of UDGs and UCDs as a function of environment.
The measure of environment we chose was the gravitational lensing mass surface
density since global tides are thought to drive the destruction of UDGs
\citep{sales2019}.
The normalized mass surface density $\kappa$, or convergence, was looked up at
the location of each galaxy using the wide-field ${{\sim}10\arcsec}$
resolution Merten maps.
The physical mass surface density $\Sigma$ was then computed by multiplying $\kappa$ by
the critical density, defined as
\begin{equation}
    \Sigma_\mathrm{crit} = \frac{c^2}{4 \pi G} \frac{D_{S}}{D_{L} D_{LS}}.
\end{equation}
The $D$'s above are angular diameter distances, where $D_S$ is the distance
to the source being lensed, $D_L$ is the distance
to the lens (cluster), and $D_{LS}$ is the distance from the lens to
the source behind it, $D_{LS} = D_S - D_L$ \citep{kneib2011}. The 
convergence maps are scaled such that $D_{LS}/D_{S} = 1$, and thus the
dependence on the redshifts of the sources used to construct the maps is
already taken into account.
In Figure \ref{fig:ucdudg-ratio}, we plot the ratio of UCDs to UDGs in bins of
mass surface density across all six clusters.
With only a small fraction of the area of the clusters exhibiting mass densities greater than 
$1.5 \times 10^9~M_\odot~\mathrm{kpc}^{-2}$, a single bin was used for mass
densities exceeding this.
The ratio of UCDs to UDGs triples in regions of mass density
greater than $\Sigma = 1\times10^9~M_\odot~\mathrm{kpc}^{-2}$.

The only cluster without an excess of UCD candidates in the cluster field
compared to the parallel field is MACS~1149.
This cluster also fails to shows a concentration of UCDs in proximity to the
BCG.
While MACS~1149 is one of the most distant FF clusters, limiting us to the
brightest UCDs, this behaviour is not seen in MACS~0717 at a nearly identical
redshift.
MACS~1149 presents with the faintest BCG and the least ICL (see Figure \ref{fig:maps}).
It is the only cluster for which the radial distribution of UDGs plateaus
towards the BCG.\footnote{The radial distribution of UDGs also plateaus
towards the centre of MACS~0717, but there the centre is not coincident with a
BCG.}
Since tidal disruption of galaxies is thought to form the ICL
\citep[e.g.][]{burke2012},
we speculate that MACS~1149 has a weaker tidal field than the other FF
clusters, permitting UDGs to survive down to lower radii.
This may then explain the low abundance of UCDs in MACS~1149 if the disruption of
UDGs (or dwarf galaxies in general) are an important formation channel for
UCDs.
If this hypothesis is confirmed, it may then be possible to use the spatial
distribution of UDGs and UCDs as tracers of the global cluster protential.

\section{Conclusions}

In this paper, we investigate the UDGs and UCDs inhabiting the six FF
clusters---Abell~2744, MACSJ0416.1$-$2403, MACSJ0717.5$+$3745,
MACSJ1149.5$+$2223, Abell~S1063 and Abell~370---and their relation to each
other and other structures present in the clusters.
The results of this paper are as follows:
\begin{enumerate}
    \item The six FF clusters are the most massive and distant
        ($0.308 < z < 0.545$) clusters in which UDGs have been found,
        with each cluster hosting between ${\sim}200$ to ${\sim}1400$ UDGs.
        The total number of UDGs in these clusters is
        consistent with the abundance halo-mass relation defined at $z \sim
        0.05$ in groups and less massive clusters.
        The slope of the relation is weakly non-linear ($N_\mathrm{UDG}
        \propto [M_{200}]^{1.13}$) at the $2\sigma$ level.
        With a slope above unity, it is possible that UDGs are more easily
        destroyed in low-mass halos and/or that UDGs may be created in
        clusters.

    \item We find that UDGs tend to not be distributed uniformly in the
        cluster core fields.
        Only in the most relaxed FF cluster, Abell~S1063, is the projected
        spatial distribution consistent with a uniform distribution.
        In at least some of the clusters, UDGs may be associated with known
        substructures late in their first infall and cluster
        merger events.

    \item The locations of UDGs and UCDs appear anti-correlated.
        UCDs are abundant in the densest environments whereas UDGs are
        deficient towards the centres of galaxy clusters.
        The ratio of UCDs to UDGs increases by roughly a factor of 3 from the
        lowest mass density regions of the clusters to the highest.
        It is interesting that MACS~1149 has the least amount of ICL and the faintest
        BCG, along with the lowest abundance of UCDs.
        Since tidal disruption of low-mass galaxies is responsible for building
        the ICL, we hypothesize that this is also responsible for producing
        UCDs.
        With many UDGs hosting compact sources, the destruction of UDGs in
        dense cluster environments may be an important formation channel of
        UCDs.
\end{enumerate}

\acknowledgements

We thank NSERC for financial support, and acknowledge support from the NSF
(AST-1616595, AST-1518294, AST-1515084 and AST-1616710).
DAF thanks the ARC for financial support via DP130100388 and DP160101608. 
AJR was supported as a Research Corporation for Science Advancement Cottrell Scholar.

Based on observations made with the NASA/ESA \textit{Hubble Space Telescope}, obtained
from the data archive at the Space Telescope Science Institute, and associated
with the Frontier Fields program.

STScI is operated by the Association of Universities for Research in
Astronomy, Inc.\ under NASA contract NAS 5-26555.

This work utilizes gravitational lensing models produced by PIs Bradač,
Natarajan \& Kneib (CATS), Merten \& Zitrin, Sharon, Williams, Keeton, Bernstein
and Diego, and the GLAFIC group. This lens modeling was partially funded by
the HST Frontier Fields program conducted by STScI. STScI is operated by the
Association of Universities for Research in Astronomy, Inc. under NASA
contract NAS 5-26555. The lens models were obtained from the Mikulski Archive
for Space Telescopes (MAST).

The scientific results reported in this article are based in part on data
obtained from the \textit{Chandra} Data Archive.

This research made use of Astropy,\footnote{\url{http://www.astropy.org}} a
community-developed core Python package for Astronomy \citep{astropy:2013,
astropy:2018}.

This research has made use of the NASA/IPAC Extragalactic Database (NED),
which is operated by the Jet Propulsion Laboratory, California Institute of
Technology, under contract with the National Aeronautics and Space
Administration.

This research has made use of NASA’s Astrophysics Data System Bibliographic
Services.

Facilities: \facility{HST (ACS, WFC3)}, \facility{CXO}

\appendix

\section{Lopsidedness of the UDG distribution in the parallel fields}
\label{sec:lopsidedness-parallel}

In Figure \ref{fig:lopsidedness-parallel}, we show the results of the same
`lopsidedness' test described in \S \ref{sec:spatialdist} and Figure
\ref{fig:lopsidedness}, but now for the six parallel fields. In contrast to
the cluster core fields, where only one cluster has a spatial distribution of
UDGs consistent with being evenly distributed around the cluster centre, UDGs
in the parallel fields are much more evenly distributed.

\begin{figure*}
	\includegraphics[width=0.50\textwidth]{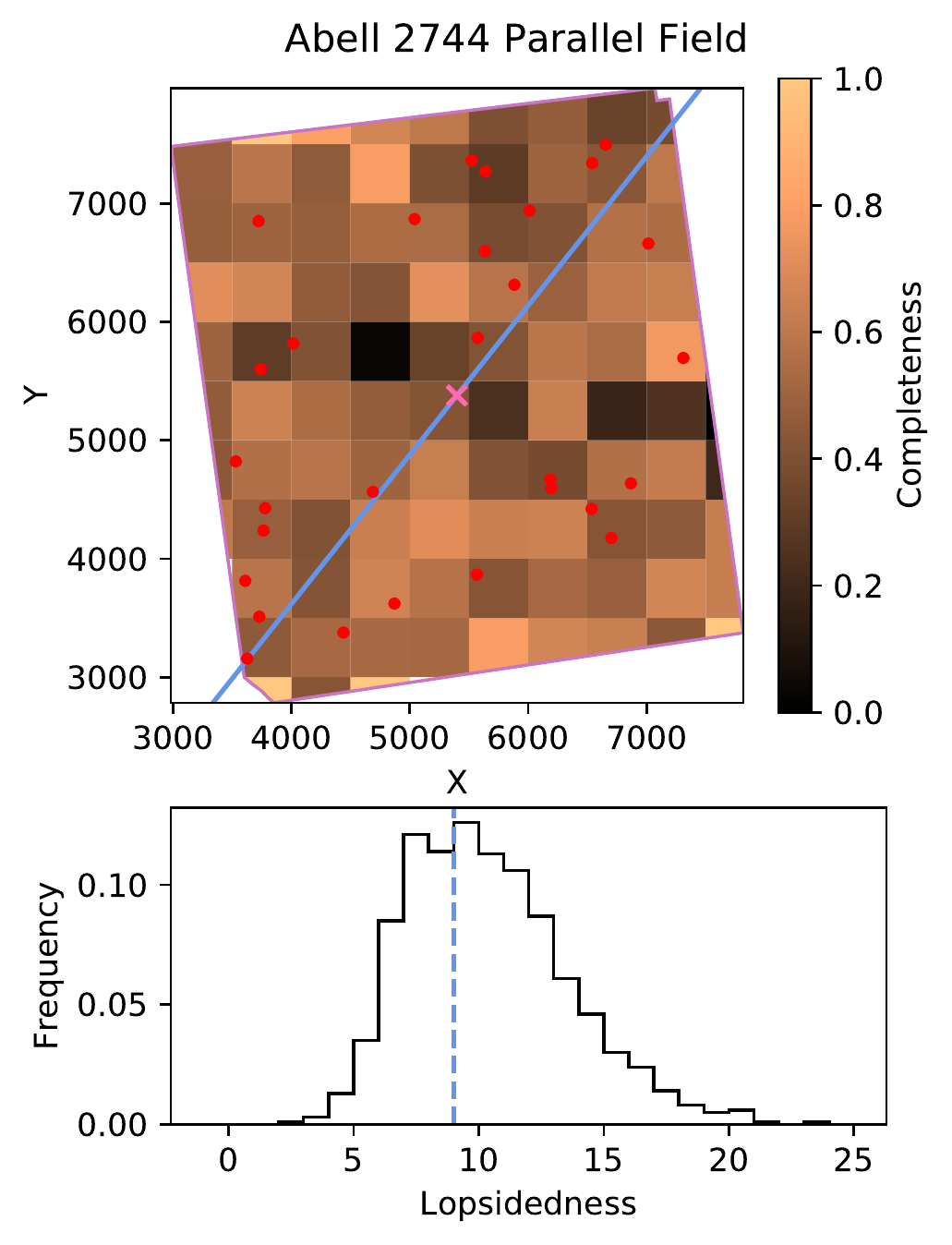}
	\includegraphics[width=0.50\textwidth]{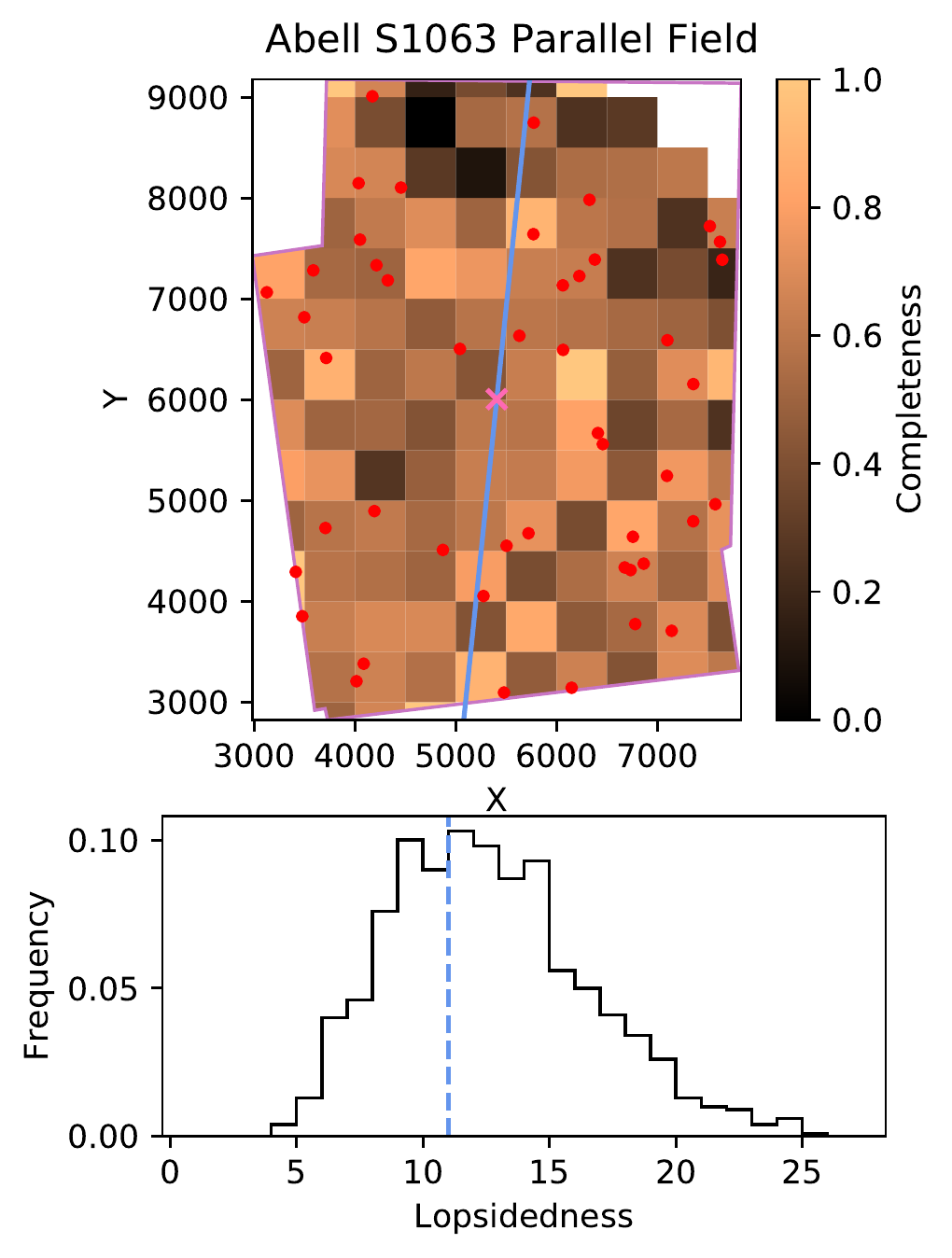}
	\caption{
    Same as Figure \ref{fig:lopsidedness} but for the parallel fields.
    \label{fig:lopsidedness-parallel}
	}
\end{figure*}
\begin{figure*}
    \ContinuedFloat
	\includegraphics[width=0.50\textwidth]{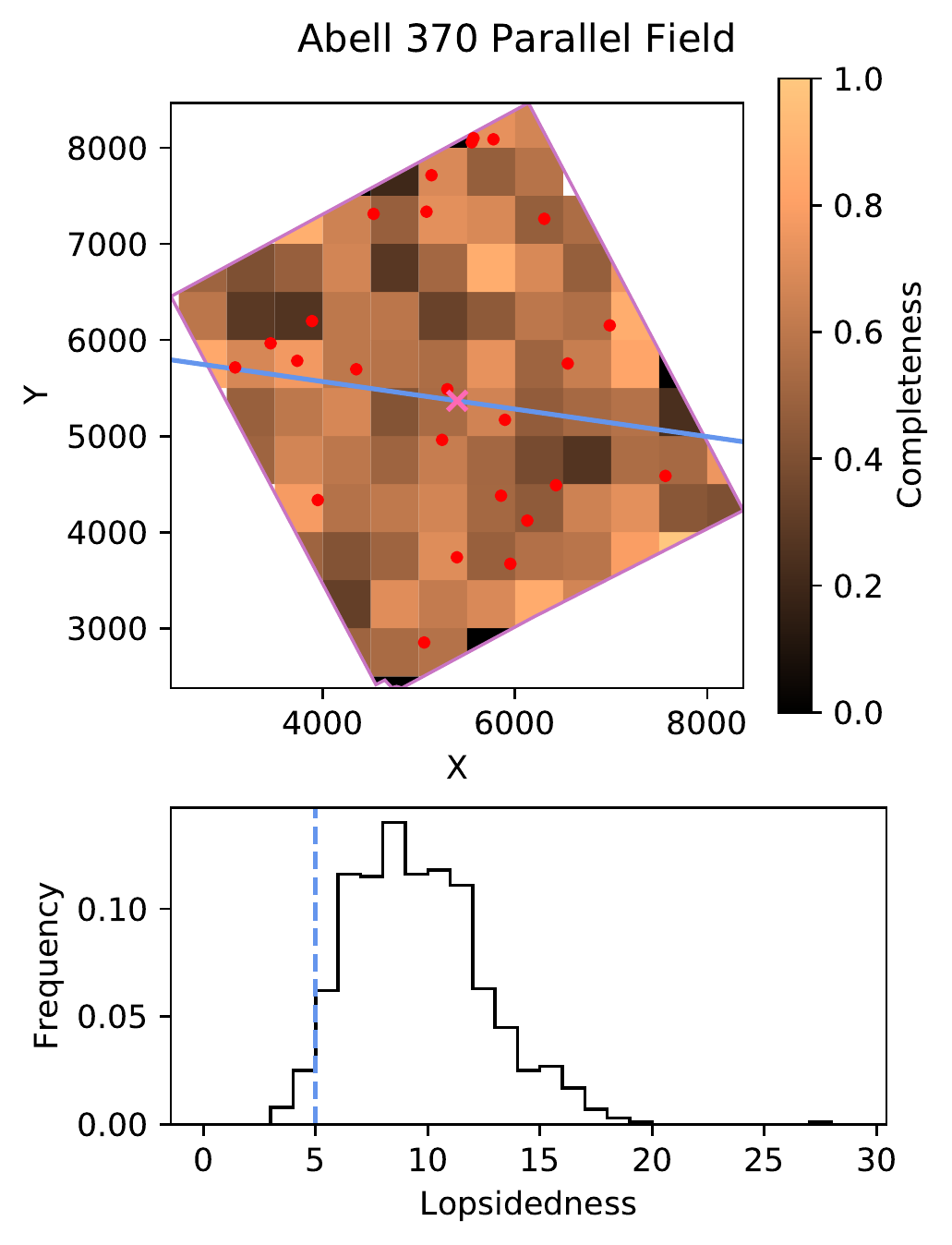}
	\includegraphics[width=0.50\textwidth]{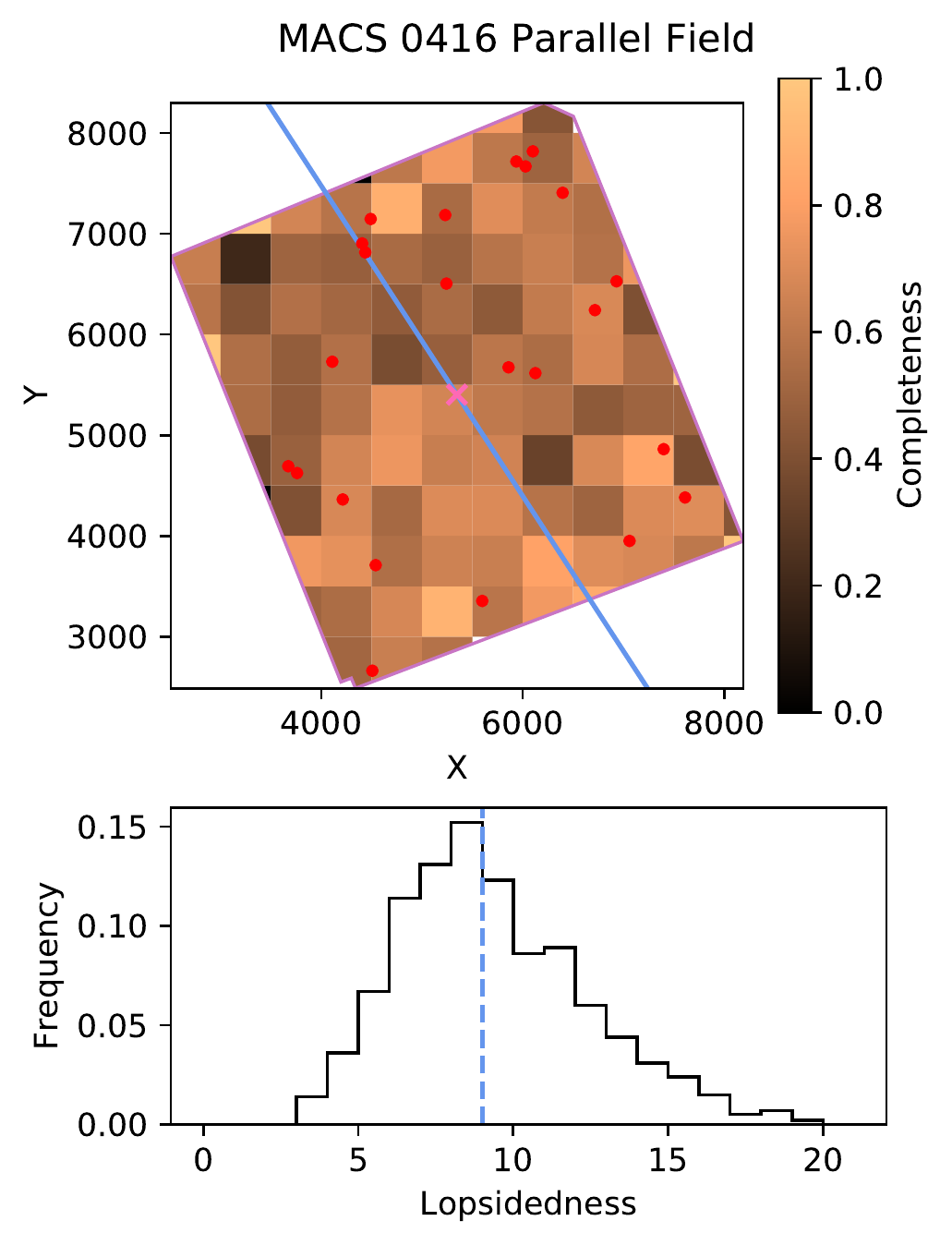}
    \caption{(Continued.)}
\end{figure*}
\begin{figure*}
    \ContinuedFloat
	\includegraphics[width=0.50\textwidth]{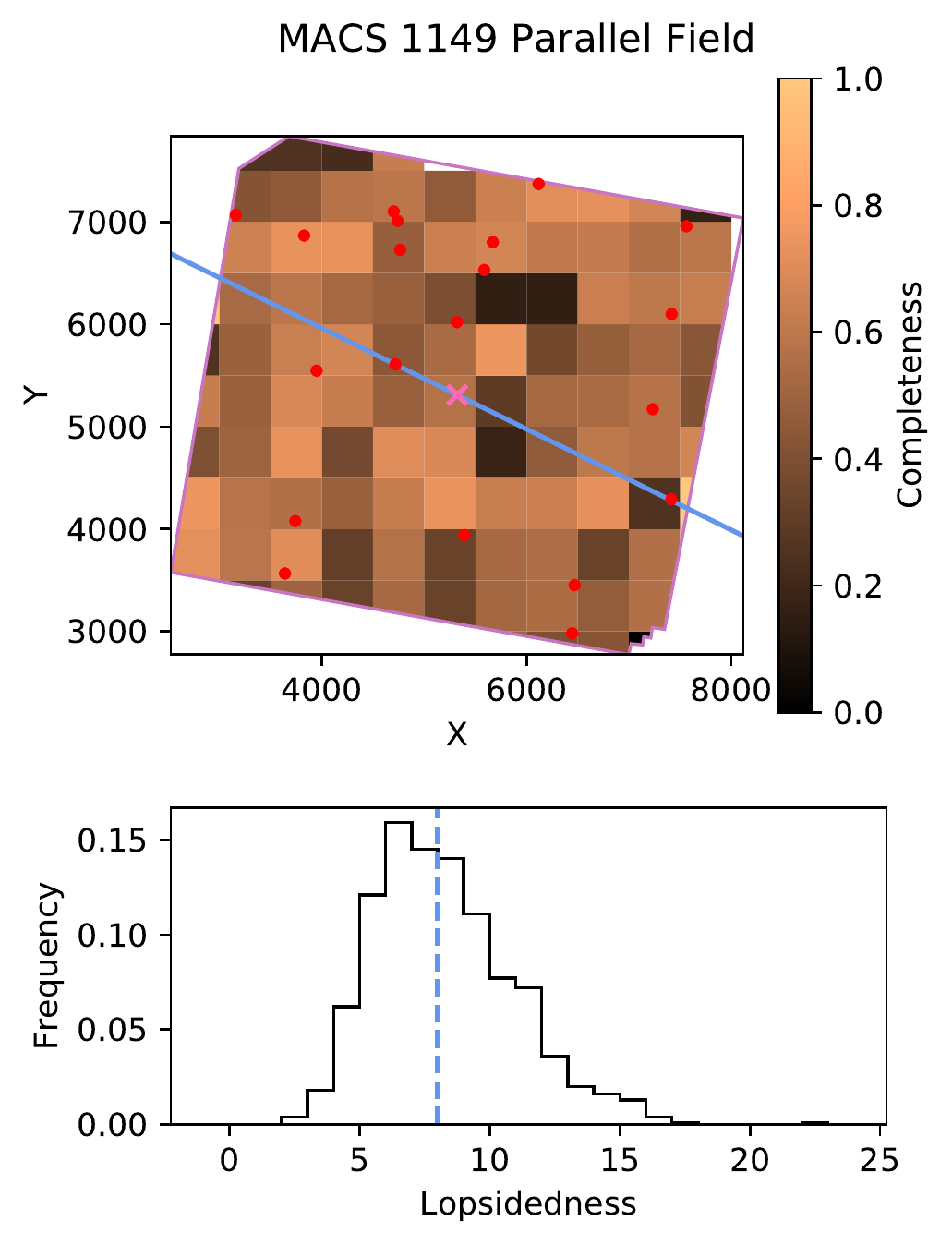}
	\includegraphics[width=0.50\textwidth]{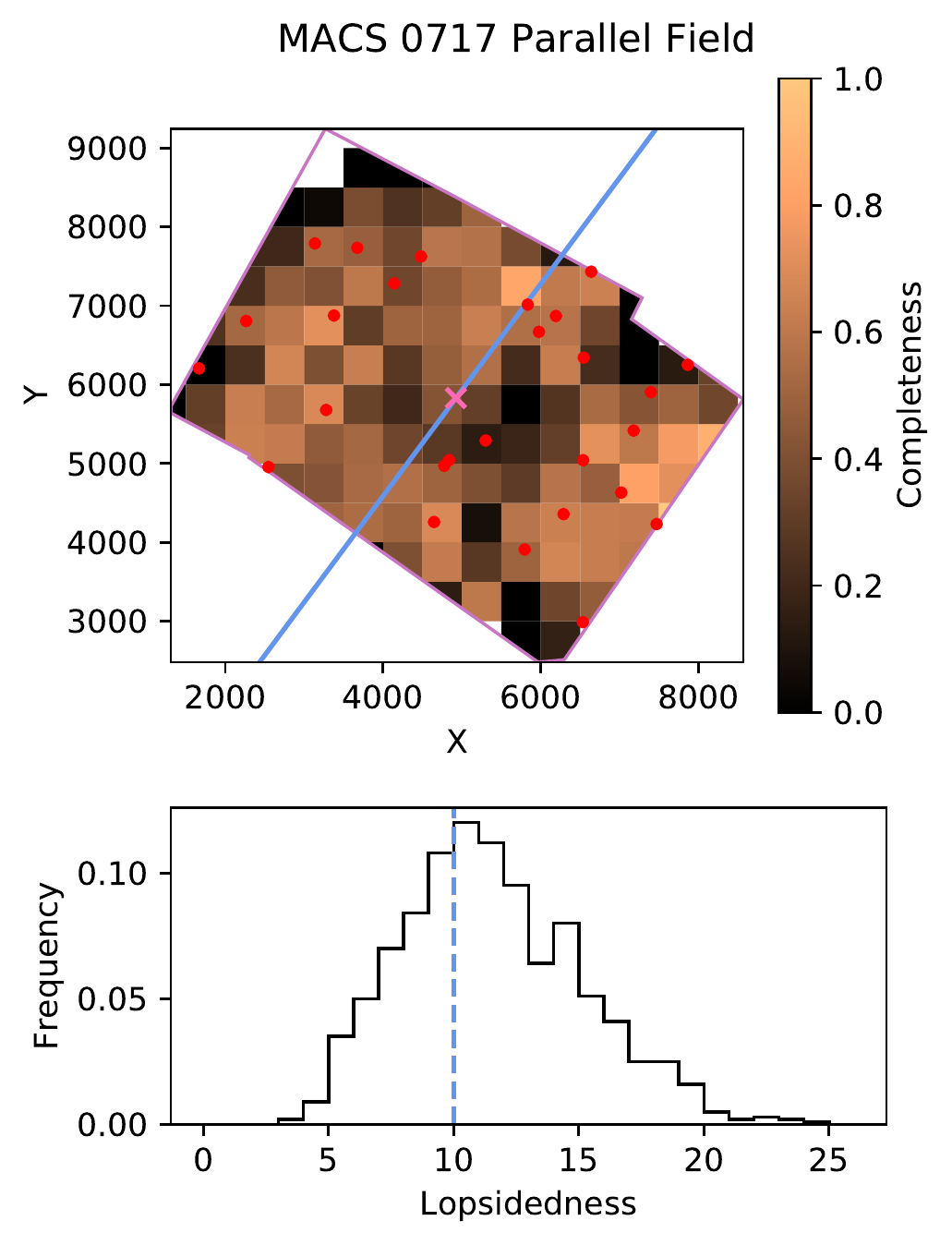}
    \caption{(Continued.)}
\end{figure*}

\section{Artificial star tests}\label{sec:artstartests}

\begin{figure*}
	\includegraphics[width=\textwidth]{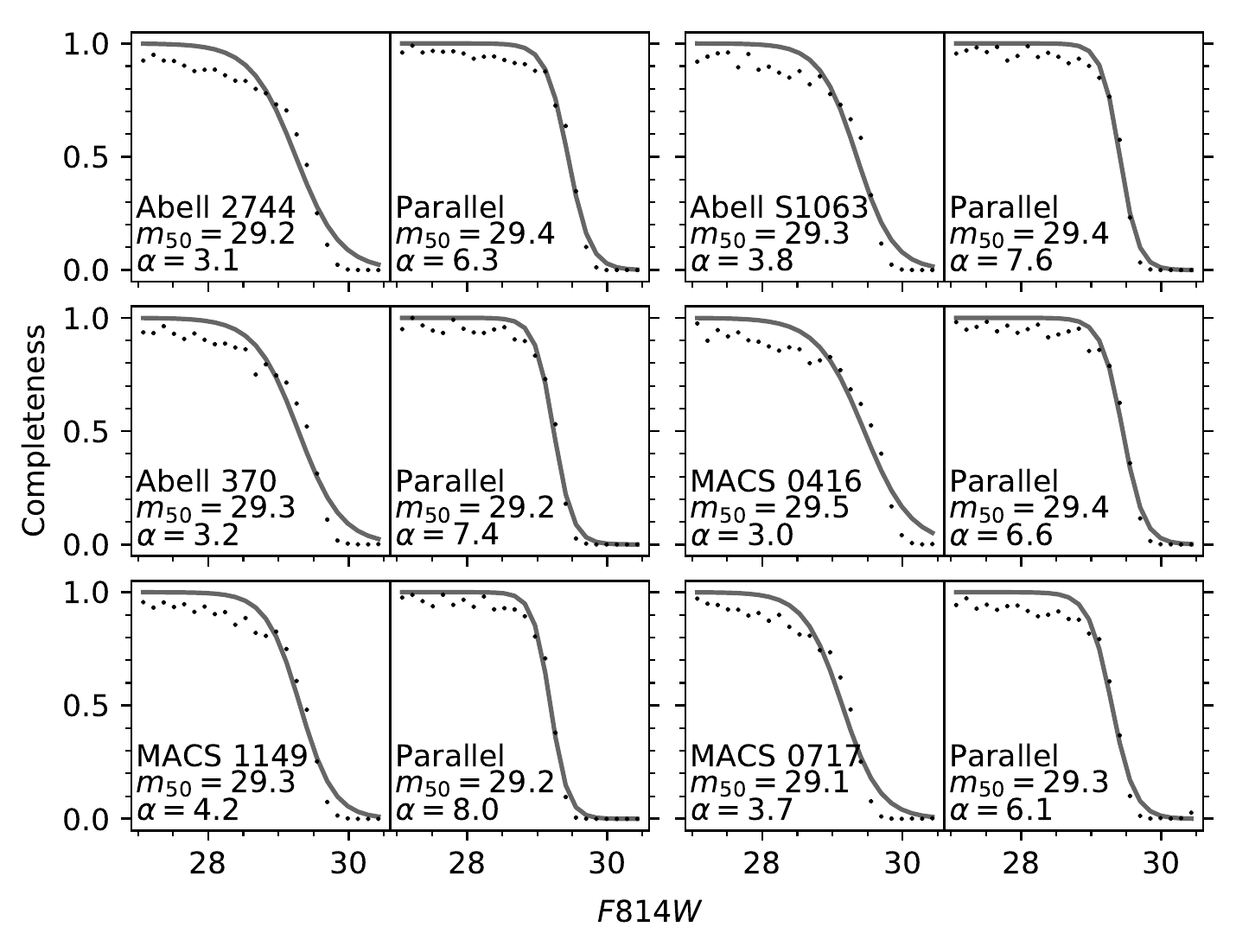}
	\caption{
    Fraction of artificial stars recovered in bins of injected magnitude for
    each of the cluster and parallel fields. The bins are 0.15 mag in size.
    The solid line is the best fit function of the form $f(m) = (1 +
    \exp(\alpha(m - m_{50}))^{-1}$, where $m_{50}$ is the magnitude at which
    50\% of the sources are recovered and $\alpha$ is the steepness at which
    the completeness drops off, both of which are listed in each panel.
    \label{fig:starcompleteness}
    }
\end{figure*}

\begin{figure*}
	\includegraphics[width=\textwidth]{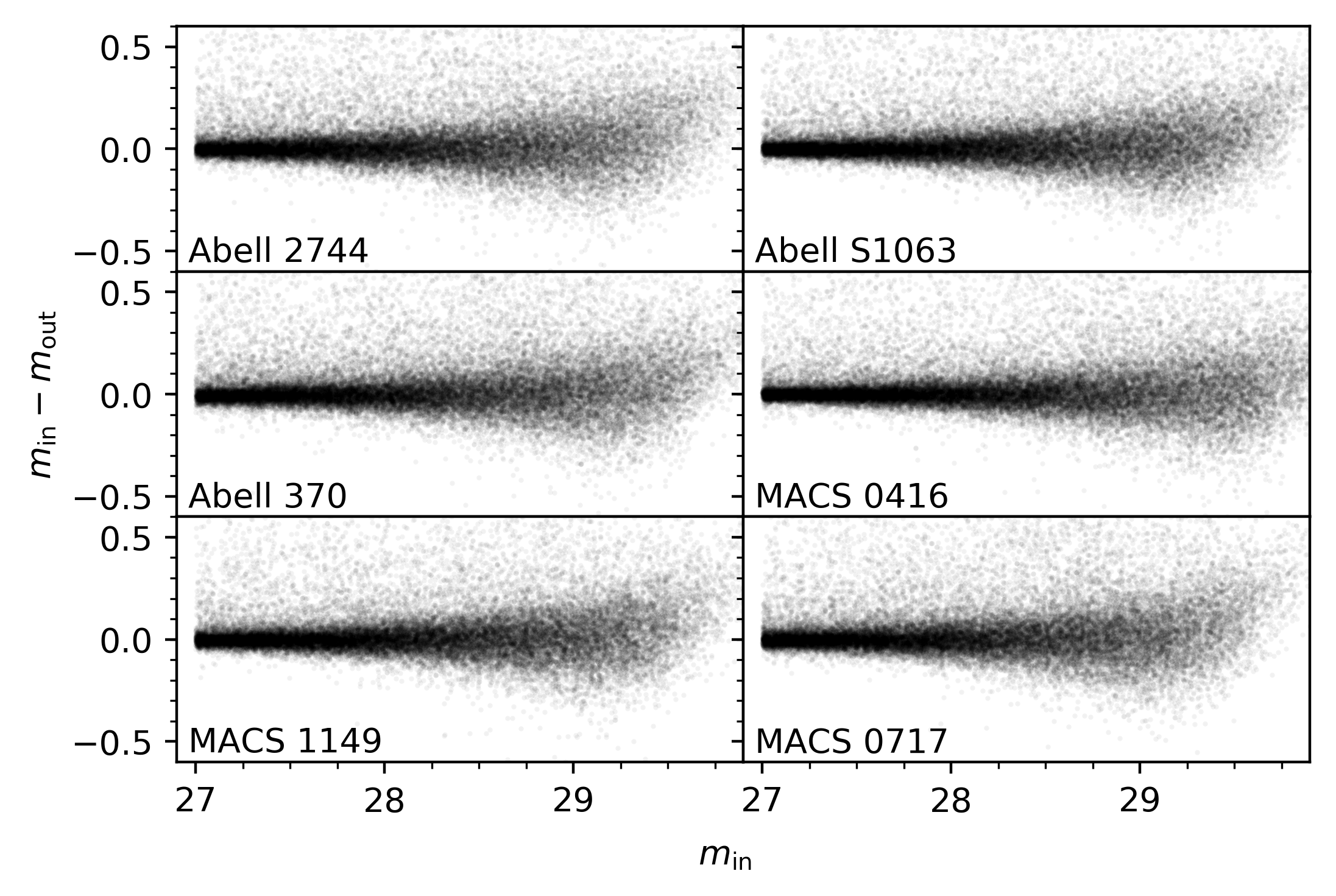}
	\caption{
    Difference between injected and recovered magnitudes for artificial stars.
    \label{fig:magerr}
    }
\end{figure*}

Artificial stars were used to test the completeness of our point source
detection.
A total of 50,000 artificial stars (in batches of 10,000) were injected into
each of the cluster and parallel field $F814W$ images at random positions with
total magnitudes chosen uniformly from the range $27 < m < 30.5$.
At each position, the \textsc{PSFEx} model was scaled to match the desired
magnitude in a 4 pixel diameter aperture. 
In Figure \ref{fig:starcompleteness}, the fraction of artificial stars
detected is plotted in bins of injected magnitude. The completeness is modeled
using the function 
\begin{equation}
f(m) = \frac{1}{1 + e^{\alpha(m - m_{50})}},
\end{equation}
where $m_{50}$ is the magnitude at which the completeness falls to 50\% and $\alpha$
determines how steep the completeness drops off \citep{harris2016}.
This simple parameterization is a much better description of the completeness
behaviour in the parallel fields than in the cluster core fields, where the
large galaxies and the ICL begin picking away at the completeness at
magnitudes well below $m_{50}$.
However, the estimate obtained for $m_{50}$ is more than adequate.
These results agree with the FF $5\sigma$ point source depths of ${\sim}29$ AB
magnitude reported by \cite{lotz2017}. 
Figure \ref{fig:magerr} shows the accuracy of recovered magnitudes.
\\

\begin{figure*}
	\includegraphics[width=\textwidth]{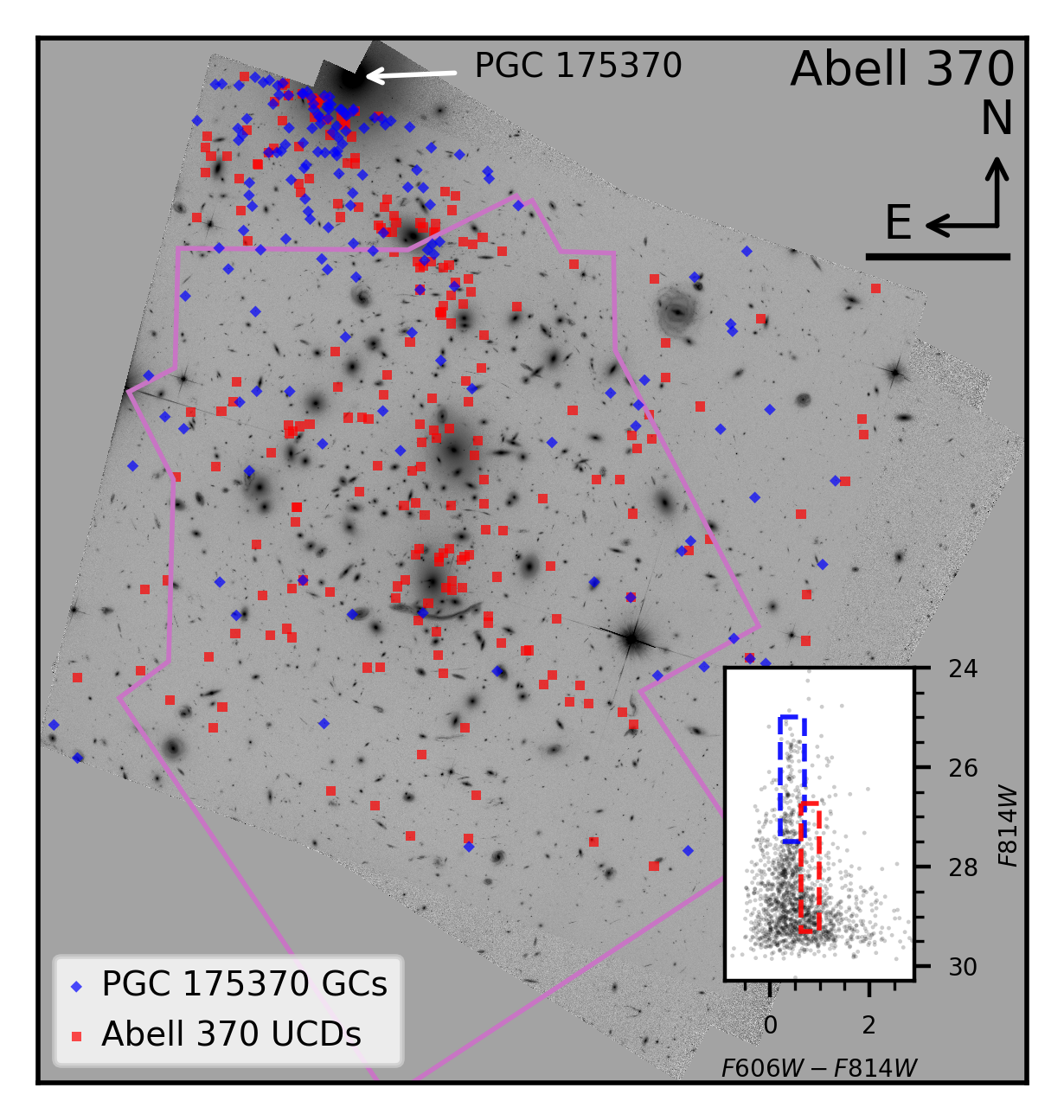}
	\caption{
    $F814W$ image of Abell~370.
    The location of PGC~175370 is marked with the arrow.
    The black line below the compass measures 200 kpc in length.
    Marked in blue are globular cluster candidates likely belonging to the
    foreground galaxy PGC~175370. 
    The foreground GC candidates are selected with the blue selection box in
    the inset CMD in the bottom right.
    Abell~370's UCD candidates are marked in red, selected using the red box
    in the CMD.
    In contrast to Figure \ref{fig:maps}, UCD selection was not limited to the
    WFC3 region, shown as the pink outline, but limiting our analysis to the
    WFC3 region removes the most likely contaminants.
    \label{fig:pgc175370}
    }
\end{figure*}

\bibliographystyle{apj}

\end{document}